\documentstyle[multicol,aps,prl,epsf,psfig]{revtex}
\begin{document}
\def\thefootnote{\fnsymbol{footnote}}
\def\kB{{k_{\scriptscriptstyle B}}} 
\def\gD{{g_{\scriptscriptstyle{\rm D}}}} 
\def\gcD{{{\rm g}_{\scriptscriptstyle{\rm D}}}} 
\def\HD{{H_{\rm D}}} 
\def\sH{{\sigma_H}}
\def\eHDkT{{{\rm e}^{-H_{\rm D}/(\kB T)}}}
\def\eHkT{{{\rm e}^{-H/(\kB T)}}} 
\def\DHv{{\Delta H_{\rm vH}}} 
\def\DHc{{\Delta H_{\rm cal}}} 
\def\DHvDHc{{\Delta H_{\rm vH}/\Delta H_{\rm cal}}}
\title{Polymer Principles of Protein Calorimetric Two-State Cooperativity}
\author{H\"useyin KAYA and Hue Sun CHAN}
\address{Department of Biochemistry, and\\
Department of Medical Genetics \& Microbiology\\
Faculty of Medicine, University of Toronto\\
Toronto, Ontario M5S 1A8, Canada}
\maketitle

\begin{abstract} 
The experimental calorimetric two-state criterion requires the van't 
Hoff enthalpy $\Delta H_{\rm vH}$ around the folding/unfolding transition 
midpoint to be equal or very close to the calorimetric enthalpy 
$\Delta H_{\rm cal}$ of the entire transition. We use an analytical model 
with experimental parameters from chymotrypsin inhibitor 2 
to elucidate the relationship among several different van't Hoff enthalpies 
used in calorimetric analyses. Under reasonable assumptions, the 
implications of these $\Delta H_{\rm vH}$'s being approximately equal 
to $\Delta H_{\rm cal}$ are equivalent: Enthalpic variations among 
denatured conformations in real proteins are much narrower than some
previous lattice-model estimates, suggesting that the energy landscape 
theory ``folding to glass transition temperature ratio'' $T_{\rm f}/T_{\rm g}$ 
may exceed 6.0 for real calorimetrically two-state proteins. 
Several popular three-dimensional lattice protein models, with different
numbers of residue types in their alphabets, are found to fall short of the
high experimental standard for being calorimetrically two-state. 
Some models postulate a 
multiple-conformation native state with substantial pre-denaturational
energetic fluctuations well below the unfolding transition temperature 
and/or predict a significant post-denaturational continuous conformational 
expansion of the denatured ensemble at temperatures well above the transition 
point. These scenarios either disagree with experiments on protein 
size and dynamics, or are inconsistent with conventional interpretation 
of calorimetric data. However, when empirical linear baseline subtractions 
are employed, the resulting $\Delta H_{\rm vH}/\Delta H_{\rm cal}$'s for
some models can be increased to values closer to unity; and baseline 
subtractions are found to correspond roughly to an operational definition 
of native-state conformational diversity. These 
results necessitate a re-assessment of theoretical models and experimental 
interpretations. 
\end{abstract}
{\phantom{\bf Key words:\ }}baseline subtraction; native state definition\\
\begin{multicols}{2}
%=========================================================================
\vfill\eject 

\centerline{\large\bf Introduction}

$\null$

In recent years, protein folding has been investigated extensively by
statistical mechanical modeling (see reviews in Refs.~1--14, Refs.~15--23, 
and references therein). The relevance of these models to the 
basic understanding of microscopic energetics is premised on the tenet that 
macroscopic properties of a system are consequences of the properties of its 
microscopic constituent parts. It follows that insight and rationalization can 
be gained by constructing models and testing whether the presumed microscopic 
interactions are effective in reproducing experimental macroscopic 
behaviors.$^{24}$ High-resolution force-field potentials have been used to 
study protein folding$^{25}$ and unfolding.$^{26-28}$ 
Obviously, atomistic models are indispensable 
for structural details. But at present it is not computationally feasible 
to use them to model thermodynamics and kinetics at millisecond or longer 
time scales. Also, it remains an open question whether empirical force 
fields would ultimately be adequate for predicting dynamics over long
simulations.$^{29}$ Currently, a significant fraction of thermodynamics and 
kinetics data of proteins can only be addressed by complementary approaches, 
mainly via polymer models with highly simplified representations of the 
geometry and interactions of the polypeptide chain.$^{1-4,15,30}$ Aside from 
their computational tractability, it is hoped that these simplified models may 
lead to the development of novel, (as-yet-undiscovered$^{31}$) concepts. 
Such ``mesoscopic'' organizing principles$^{31}$ may be needed to bridge our 
understanding over gaps of many orders of magnitude in time and length 
scales separating the fundamental constituent atomic processes and the 
global features of a bio-macromolecule.\\ 

\noindent
{\bf Simple self-contained polymer models can be used to explore microscopic
energetics of proteins.}\\

How do simple polymer protein models contribute to our physical understanding
of proteins? Typically, the ingredients of such a model are (i) a 
conformational space that accounts for chain connectivity and excluded 
volume, and is sufficiently simple to be enumerated exhaustively$^{4,32}$ 
or sampled extensively,$^{3,7,11}$ and (ii) a set of rules 
(a potential function) that describes the ``microscopic'' interactions 
among the constituent parts of the chain. The most important feature of such 
a model is the conceptual clarity it offers because it is self-contained. 
This means that all properties and predictions of the model are derived 
solely from the postulated elementary microscopic ingredients. In particular, 
conformational ensembles are determined by applying the model potential 
function (ii) to ascertain the energetic favorability of every conformation 
in the model conformational space (i). Most recent lattice protein 
models belong to this category. However, some protein models are not 
self-contained in this sense. In some thermodynamic treatments$^{33-36}$, for 
example, the unfolded or denatured state of a protein is postulated to contain 
only random-coil-like conformations, but with no specification as to what 
microscopic interactions are responsible for such a remarkable property in
discriminating against compact nonnative conformations. (See discussions 
in Ref.~23.) As such, non-self-contained models involve either unspecified or
unjustified mechanisms that are not explicitly considered as parts of their 
microscopic potential functions. Therefore, their explanatory power is 
limited because they cannot make a full logical connection between the 
macroscopic properties they predict and the microscopic interactions they 
explicitly consider, though they can provide important insight and be very
useful in other respects.

Self-contained simple polymer models of proteins help frame our discourse in 
terms of basic physical interactions. They sharpen our focus on whether certain 
global properties can or cannot arise from the microscopic interactions 
presumed by a model. In these models, however, the necessity to simplify 
implies that one has to rely to a degree on intuitive judgement in the design 
of appropriate model representations 
to capture polypeptide properties. In principle, many simple models can 
give similar results. A successful predictions can therefore be fortuitous. It 
follows that the ability to reproduce a protein property is necessary but 
not sufficient for the validity of the presumed microscopic features of a 
model. On the other hand, if properties of a model are in disagreement with 
experimental data, it is a clear indication of deficiencies.
Since simple models appear to enjoy a high degree of latitude in their
design, it might be expected that reproducing general, ``generic''$^{37}$ 
properties of proteins would be straightforward. This is not the case.
To the contrary, using simple models with physically plausible interactions 
to reproduce several thermodynamic$^{23,38}$ and kinetic$^{19,39}$ 
properties of proteins has been shown to be not trivial and requires in-depth 
analyses. This may be a blessing in disguise, because it means that a lot 
can be learnt about microscopic protein energetics from generic protein 
properties by using the latter as restrictive experimental constraints on 
models, to provide insight into what form of microscopic interactions 
are more likely to be proteinlike.\\

\noindent
{\bf The calorimetric criterion for thermodynamic two-state 
cooperativity requires a narrow denatured-state enthalpy distribution.}\\

One generic protein property that apparently has not been fully appreciated
by modelers is the calorimetric two-state behavior of many small 
single-domain proteins,$^{40,41}$ which requires that the van't Hoff enthalpy 
$\Delta H_{\rm vH}$ 
around the folding/unfolding transition midpoint to be equal or very close to 
the calorimetric enthalpy $\Delta H_{\rm cal}$ of the entire transition. 
Thermodynamic properties of several simple polymer models have recently
been compared with this experimental criterion for two-state 
cooperativity.$^{22,23}$ One of us$^{23}$ argued that, under reasonable 
assumptions, the calorimetric two-state condition requires the average 
enthalpy difference between the denatured and native ensembles around the 
heat denaturation midpoint not to further increase appreciably as the 
temperature is raised to complete the unfolding process. From 
analyses of analytic as well as two-dimensional lattice models, this is found 
to imply that the enthalpy distribution among the denatured ensemble of 
conformations has to be narrow in comparison with the average enthalpy 
difference between the native state and the denatured state.$^{23}$ In the 
present study, we provide further support for this view by determining
systematically the effects of using several slightly different common 
definitions of van't Hoff enthalpy for the calorimetric two-state criterion. 

A number of two-dimensional lattice protein models have been evaluated 
against the calorimetric criterion.$^{23}$ Interestingly and unexpectedly, 
both a G\=o$^{15,19}$ and a G\=o-like HP+ (Ref.~19) model are found to be
far away from being calorimetrically two-state. Apparently, insofar as the 
underlying chain model is highly flexible, even for these models with 
native-specific pairwise additive contact interactions (these interaction
schemes are sometimes referred to as being ``nearly maximally 
unfrustrated''$^{42,43}$), the denatured enthalpy distributions in these
two-dimensional models are still too board to satisfy the calorimetric 
two-state standard. Based on these results, it has been suggested that a 
cooperative interplay between local and nonlocal interactions in proteins may 
be necessary to give rise to calorimetrically two-state behaviors.$^{23}$ In 
the present work, we evaluate six three-dimensional lattice protein models.
These include two-$^{44}$ and three-letter$^{45}$ models, a G\=o model,$^{46}$ 
a ``solvation'' model$^{47}$ and 20-letter models with$^{48}$ and 
without$^{49}$ sidechains. Their thermodynamics are checked against the 
calorimetric criterion. We also evaluate the physical pictures of native 
and denatured states offered by some of these models in light of other 
experimental measurements on protein folding/denaturation transitions.\\

\centerline{\large\bf Results and Discussion}

$\null$ 

\noindent
{\bf Overview of an analytical treatment.}\\

To provide a basic theoretical underpinning, we
first re-examine several definitions of van't Hoff enthalpy 
($\Delta H_{\rm vH}$'s) in the protein folding literature, and the
consequences of using different $\Delta H_{\rm vH}$'s in the calorimetric
two-state criterion $\Delta H_{\rm vH}/\Delta H_{\rm cal}\approx 1$. 
The main result of this section, to be demonstrated below, is that under 
reasonable, minimal assumptions regarding protein conformational properties, 
calorimetric two-state criteria using several commonly employed 
$\Delta H_{\rm vH}$'s imply essentially equivalent requirements on a 
protein's density of states.$^{23}$ 
We approach this by comparing the $\Delta H_{\rm vH}/\Delta H_{\rm cal}$ 
values using different $\Delta H_{\rm vH}$'s computed for a series of 
analytical models with a wide range of thermodynamic cooperativities. 

We begin by recalling a few basic relations. As discussed in detail 
previously,$^{23}$ the main thermodynamic quantities of interest for the 
issues at hand are the excess enthalpy and heat capacity. Experimentally, 
raw calorimetric data consists of heat capacity scans over a range of 
temperatures, from which an excess enthalpy
$$
\langle\Delta H(T)\rangle = 
\langle H(T)\rangle - H_{\rm N}\;
\eqno(1)
$$
as a function of absolute temperature $T$ can be obtained by standard 
baseline subtraction and numerical integration techniques.$^{41}$ Here
$H$ is the enthalpy of the entire ``excess'' system,$^{23,41}$ $H_{\rm N}$
is the enthalpy of the native state, and $\langle\dots\rangle$ denotes 
Boltzmann averaging. In general, the native enthalpy $H_{\rm N}$ should be 
replaced by a Boltzmann average $\langle H_{\rm N}(T)\rangle$ over 
conformational variations in the native state. (See discussions below on
20-letter models with and without sidechains.) Here we adopt as a working
assumption that the native state become effectively a single conformation
with a single temperature-independent enthalpy value after proper baseline 
subtractions.$^{23}$ The calorimetric enthalpy $\Delta H_{\rm cal}$ $=$
$\langle\Delta H(T_1)\rangle$ at a sufficiently high temperature $T_1$ at
which the heat denaturation process is completed ($T_1$ may be formally
taken to be $\infty$ in model considerations).$^{23}$ The expression for 
the excess heat capacity function
$$
C_P = {\frac {\partial\langle\Delta H(T)\rangle} {\partial T}}
= {\frac {\langle H^2(T)\rangle-\langle H(T)\rangle^2} {\kB T^2}} \; ,
\eqno(2)
$$
follows from standard statistical mechanics,$^{23}$ where $\kB$ is 
Boltzmann's constant. Equation~(2) corresponds to $\Delta C_P$ in 
the calorimetric literature ($\Delta C_{P,{\rm tr}}$ in 
Ref.~41). We drop the symbol $\Delta$ here for the excess heat capacity
as in Ref.~23 to simplify notation.

Several different definitions of $\DHv$ have been put forth in
the protein calorimetric literature.$^{22,23,40,41,50,51}$ In general,
their values can be very different. This raises the possibility of 
complications in comparison between theory and experiment.
In Ref.~23, one of us noted that while different $\DHv$'s may be different 
when the transition is far from being calorimetrically two state --- i.e.,
two-state as defined by the condition 
$\Delta H_{\rm vH}/\Delta H_{\rm cal}\approx 1$ using any one of the 
$\Delta H_{\rm vH}$'s, a semi-quantitative argument
can infer that for proteins which can be fully denatured by
heat, $\Delta H_{\rm vH}$ $\approx$ $\Delta H_{\rm cal}$ for one
$\DHv$ would imply that the same approximate equality also holds for
other $\DHv$'s. Here we substantiate this inference by quantitatively 
analyzing a class of models for protein densities of states.\\

\noindent
{\bf Definitions of protein folding van't Hoff enthalpies.}\\

In general, a temperature-dependent van't Hoff enthalpy is given by
$$
\Delta H_{\rm vH}(T) = \kB T^2 {\frac {d\ln K^{\rm eff}} {dT}}
= \kB T^2 {\frac {1} {\theta(1-\theta)}} {\frac {d\theta} {dT}} \; ,
\eqno(3)
$$
where $K^{\rm eff}$ is the apparent$^{52,53}$ or effective$^{22,51}$
equilibrium constant of the system, and $\theta$ $=$ $\theta(T)$ is a 
two-state progress parameter for tracking the transition process;
$K^{\rm eff}=\theta/(1-\theta)$ and $\theta$ takes values from 
zero (at low temperatures in the present cases) to unity 
(at high temperatures).  For heat denaturation of proteins, 
$\theta=0$ and $\theta=1$ correspond respectively to the completely native 
(fully folded) and fully denatured (unfolded) states.\footnote{$\theta$ 
is equivalent to Lumry et al.'s$^{52}$ 
($[\langle\alpha\rangle(T)-\langle\alpha\rangle_{\rm A}(T)]$ $/$
$[\langle\alpha\rangle_{\rm B}(T)-\langle\alpha\rangle_{\rm A}(T)]$,
where $\alpha$ is an observable [their Eq.~(4)].}
Therefore, at the midpoint temperature $T_{\rm midpoint}$ of the 
parameter $\theta$, i.e., when $\theta(T=T_{\rm midpoint})=1/2$,
$$
\Delta H_{\rm vH} = 4\kB T_{\rm midpoint}^2 {\frac {d\theta} {dT}} 
\biggr\vert_{T=T_{\rm midpoint}} \; .
\eqno(4)
$$
As in Ref.~23 and is customary in the calorimetric literature, 
$\Delta H_{\rm vH}$ is understood to be evaluated at a certain midpoint 
temperature when its $T$ dependence is not shown explicity. 

It follows that different choices of $\theta$ would result in different
van't Hoff enthalpies and different midpoint temperatures. The theoretical
population-based $\DHv$ in Ref.~23 corresponds to $\theta$ = [D] ---
the denatured fraction of the total population, and a midpoint temperature
$T_{1/2}$ at which one half of the chain population is denatured.
Here we use $\kappa_0$ to denote the $\DHv/\DHc$ ratio of this 
population-based van't Hoff enthalpy to the calorimetric enthalpy.
Experimentally, the heat absorbed by the system is often used to quantitate
the degree of progress of the transition process under a two-state 
assumption by setting $\theta=\langle\Delta H\rangle/\Delta H_{\rm cal}$, 
with a corresponding midpoint temperature $T_d$ at which
one half of the total calorimetric heat ($\Delta H_{\rm cal}/2$) has 
been absorbed (Ref.~51). This leads to a van't Hoff enthalpy which is
proportional to the excess specific heat at $T_d$ (see below).

On the other hand, a ``square-root'' van't Hoff enthalpy formula has also
been used by Privalov and coworkers$^{40,50}$ to analyze experimental data. 
It takes the form
$$
\Delta H_{\rm vH} = 2T_{\rm midpoint}\sqrt{\kB C_P(T_{\rm midpoint})} \; .
\eqno(5)
$$
Apparently, this corresponds to setting $\theta(T)$ $=$
$\langle\Delta H(T)\rangle/\DHv$, and assuming that it is a valid
progress parameter.
Equation~(5) is used in conjunction with 
either the peak temperature $T_{\rm max}$ of $C_P$ (Ref.~40) or $T_d$ (Ref.~50)
as midpoint temperatures at which $\theta=1/2$ is presumably a
good approximation (see also Ref.~23). To ascertain the effects of 
different $\DHv$'s on the calorimetric criterion, we compare the 
population-based $\kappa_0$ defined above with the following 
possible van't Hoff to calorimetric enthalpy ratios using different
midpoint temperatures for the square-root formula$^{23,40,50}$: 
\begin{eqnarray*}
%$$
%\eqalign{
\kappa_1 & = & {2T_{1/2}\sqrt{\kB C_P(T_{1/2})}}/{\DHc} \; , \\
\kappa_2 & = & {2T_{\rm max}\sqrt{\kB C_P(T_{\rm max})}}/{\DHc} \;
,\;\;\;\;\;\;\;\;\;\;\;\;\;\;\;\;\;\;\;\;\;\;\;\; (6)\\
\kappa_3 & = & {2T_{d}\sqrt{\kB C_P(T_{d})}}/{\DHc} \; . \\
%}
%\eqno(6)
%$$
\end{eqnarray*}
%$$
%\eqalign{
%\kappa_1 & = {2T_{1/2}\sqrt{\kB C_P(T_{1/2})}}/{\DHc} \; , \cr
%\kappa_2 & = {2T_{\rm max}\sqrt{\kB C_P(T_{\rm max})}}/{\DHc} \; , \cr
%\kappa_3 & = {2T_{d}\sqrt{\kB C_P(T_{d})}}/{\DHc} \; . \cr}
%\eqno(6)
%$$
Finally, it is not difficult to see that the van't Hoff to calorimetric 
enthalpy ratio for $\theta=\langle\Delta H\rangle/\Delta H_{\rm cal}$ above 
is given$^{51}$ by $(\kappa_3)^2$. So we also consider 
$(\kappa_1)^2$ $(\kappa_2)^2$ and $(\kappa_3)^2$ as possible van't Hoff 
to calorimetric enthalpy ratios. The definitions and usage of these 
quantities are summarized in Table~I.\\

\noindent
{\bf Despite their different definitions, several van't Hoff enthalpies give
 essentially the same calorimetric two-state criterion.}\\

We now compute these different van't Hoff to calorimetric enthalpy ratios 
for a class of models that intuitively capture the most basic features 
of protein energetics, which are an essentially unique native state
as the lowest (ground) enthalpic state of the system, and a huge number 
of unfolded (denatured) conformations with higher enthalpies. For this purpose, 
we use simple random-energy-like models with Gaussian enthalpy distributions 
for the denatured states. Their (continuum) densities of states ${\rm g}(H)$ 
are given by$^{23}$ 
$$
{\rm g}(H) = \delta(H) + \theta(H){\frac {\gcD} {\sqrt{2\pi}\sH}}\;
{\rm e}^{-(H-\HD)/(2\sH^2)} \; ,
\eqno(7)
$$
where $\delta(H)$ is the Dirac delta function, the native enthalpy
$H_{\rm N}=0$, the step function $\theta(H)=1$ for $H\ge 0$, and
$\theta(H)=0$ for $H<0$. $\gcD$ ($\gg 1$) and $\HD$ are
respectively the total number and average enthalpy of the denatured
conformations, whereas the standard deviation $\sH$ specifies the
width of the enthalpy distribution among them (Figure~1); see Ref.~23 
for details. The corresponding partition function $Q=Q_{\rm N}+Q_{\rm D}$, 
whose native part $Q_{\rm N}=1$ is the statistical weight of the native
state, and the denatured part
$$
Q_{\rm D}(T)={\frac {\gcD} {\sqrt{2\pi}\sH}}\int dH\;
{\rm e}^{-(H-\HD)/(2\sH^2)} \; {\rm e}^{-H/(\kB T)} \; .
\eqno(8)
$$
Hence [D] = $Q_{\rm D}/Q$. We perform numerical integrations over 
$H$ to obtain thermodynamic averages such as native and denatured populations 
[Eq.~(8)], average enthalpy, and heat capacity as functions of temperature, 
from which the midpoint temperatures and $\kappa$'s defined above are 
determined. To simplify these calculations, rather than integrating 
through $H\rightarrow +\infty$, we use a high $H$ cutoff that 
set ${\rm g}(H)=0$ for $H>4\HD$ in Eq.~(7). The special case of a strictly 
two-state model (corresponding to $\sH\rightarrow 0$) is discussed in 
the Appendix.

For the class of models we study, we fix both the average enthalpy ($\HD$)
and entropy (parametrized by $\gcD$) of the denatured state. This leads$^{23}$ 
to an essentially constant $\DHc$ $=$ $\HD$. Only the denatured enthalpy 
distribution width $\sH$ is varied. Here we use $\HD/\kB=3\times 10^4$ 
(equivalent to $\HD= 60.0$ kcal mol$^{-1}$), and $\gcD=5.68\times 10^{38}$ 
(Figure~1). These values are the same as those used in 
our previous study.$^{23}$ They correspond approximately to the 
experimental data obtained by Jackson et al.$^{54}$ for the 
Ile$\rightarrow$Val76 mutant of chymotrypsin inhibitor 2 (CI2; see Fig.~3 
of Ref.~54). Hence we believe that realistic protein energetics can be 
explored using this class of models. 

Figure~2 shows how the model midpoint temperatures and thermodynamic 
cooperativity vary with $\sH$. The calorimetric two-state criterion allows 
for some tolerance. This is because even small single-domain proteins 
deviate slightly from a {\it strictly} two-state description,$^{33}$ 
with $\DHv/\DHc$ slightly less than unity. So we do not have to require 
model $\DHv/\DHc$ to be exactly equal to unity. Nonetheless, it 
is also clear that the experimental observation of $\DHv/\DHc\approx 1$ 
imposes severe constraints on enthalpy distributions in proteins. 
Experimentally, $\DHv/\DHc=0.96$ is reported by Fersht and coworkers$^{54}$
for CI2, other calorimetric two-state proteins have similar
$\DHv/\DHc$'s (Ref.~33.) For the present models, if the $\DHv/\DHc$'s are to 
be $\ge 0.96$, it requires $\sH\le 775$ (Figure~2b, in units of $\kB$). This 
means a very narrow denatured enthalpy distribution, as the standard 
deviation $\sH$ has to be less than or equal to 
$775/(3\times 10^4)\approx 1/40$ of the average 
enthalpic separation between the native and the denatured states, $\DHc$
(see Figure~1). Within this class of models, thermodynamic stability 
correlates with cooperativity (Figure~2a). For $\DHv/\DHc\approx 1$, the 
folding transition temperature $\approx 65^\circ$C corresponds to that 
observed experimentally.$^{54}$ However, stability decreases as the denatured 
enthalpy distribution widens. The transition temperature falls below 
$0^\circ$C when $\sH$ exceeds $\approx 1/17$ of $\DHc$. 

Figure~2a shows the relation among the three midpoint temperatures.
They are essentially identical when the model protein is 
highly cooperative (small $\sH$). The difference between $T_d$ and the other 
two temperatures increases as cooperativity diminishes. This is because when
the enthalpy distribution in the denatured state is wide (large $\sH$), 
there are more low-lying nonnative enthalpies, which tend to lower
the overall average enthalpy. As a result, more than half of the chain 
population has to be denatured (hence a higher temperature than $T_{1/2}$ 
is required) to achieve an average enthalpy of $\DHc/2$ than when the 
denatured enthalpy distribution is narrower (smaller $\sH$). This
accounts for the differences among the three $\kappa$'s [Eq.~(6)] and
$(\kappa)^2$'s in Figures~2c and d. For real two-state proteins, $T_d$ 
can differ from $T_{\rm max}$ by $\sim 1^\circ$C (Ref.~50).  On the other 
hand, $T_{\rm max}$ is practically identical to $T_{1/2}$ for a much wider 
range of cooperativity for these models. It appears that 
$T_{\rm max}\approx T_{1/2}$ is a 
consequence of $\gcD\gg 1$. Model proteins with less conformational 
freedom$^{23}$ than those considered in Figures~1 and 2 have non-negligible 
differences between $T_{1/2}$ and $T_{\rm max}$ (see Appendix and
discussions on three-dimensional lattice models below).

Figures~2c and d compare the population-based$^{23}$ $\kappa_0=\DHv/\DHc$
with experimental formulas and their variations. For this class of models,
$\kappa_0$ $=$ $\kappa_1$ $=$ $\kappa_2$ holds almost exactly. Owing to
the behavior of $T_d$ discussed above, $\kappa_3$ deviates from the other 
three $\kappa$'s when the model is not cooperative, but all four $\kappa$'s
are practically identical if their values are $\ge 0.9$.
When the enthalpy ratios $\kappa$'s are less than one, naturally the
square-root ($\kappa$) formulas Eq.~(6) gives larger van't Hoff to calorimetric
enthalpy ratios than the $(\kappa)^2$ formulas. The latter equate $\DHv$ with
$4\kB T^2_{\rm midpoint}C_P(T_{\rm midpoint})/\DHc$ (Ref.~22,40,41,51).
However, when any one of the $\DHv/\DHc$'s equals unity, it implies that 
all other $\DHv/\DHc$'s also equal unity.

These observations suggest that the following general conclusion should
be valid: Insofar as a protein can be fully denatured by heat$^{23}$
(as these models are), which implies that it has a sufficiently high 
denatured-state entropy relative to the native state (which should be 
satisfied by all proteins because of their polymeric nature), all of the 
$\DHv/\DHc$'s considered in this paper provide essentially the same 
calorimetric two-state conditions, and thus have the same requirement 
on the density of states of the proteins. 

Recently, Zhou et al.$^{22}$ used a homopolymer tetramer model
to show that it is possible to have $(\kappa_3)^2>1$,
and that the deviation from the calorimetric criterion is not simply related
to the population with intermediate enthalpies. Remarkably, the
thermodynamic properties of their continuum tetramer model are very
similar$^{23}$ to that of a lattice tetramer toy model introduced previously
by Dill et al.$^{4}$ Since the ground-state populations of these small
systems are substantial$^{23}$ even under athermal conditions ($T=\infty$),
they cannot be fully ``denatured.'' Hence this interesting and important
observation of Zhou et al. is not inconsistent with our general 
conclusion regarding proteins. The present study does not address
the application of van't Hoff analysis to chemical reactions in 
solutions$^{55}$ because of fundamental differences between chemical reactions
and the conformational transition of polymeric systems treated here.\\

%\vfill\eject

\noindent
{\bf Calorimetric two-state cooperativity implies a very low ``glass
transition'' temperature for the folding of two-state proteins.}\\

The above thermodynamic results are relevant to folding kinetics,
especially landscape theories that utilize the spin-glass approach 
put forth in the seminal work of Bryngelson and Wolynes.$^{56,57}$
It has been argued, and has been generally accepted, that in order 
for a protein to fold in a kinetically efficient manner, 
its folding transition temperature $T_{\rm f}$ must be significantly 
greater than a glass temperature $T_{\rm g}$ that characterizes the 
onset of sluggish folding kinetics as the temperature is lowered$^{58}$ 
(reviewed in Refs.~3, 4). Subsequently, based on a series of insightful
studies by Onuchic, Wolynes and coworkers,$^{45,59,60}$ it has been further
argued that a ``law of corresponding states''$^{6,59,60}$ can be used
to predict the ratio $T_{\rm f}/T_{\rm g}$ for real proteins from simulations
of a 27mer 3-letter code (3LC) model protein configured on three-dimensional 
cubic lattices$^{45,59}$ (see discussion below). This approach provided
an estimate of $T_{\rm f}/T_{\rm g}$ $=$ $1.6$ for small $\alpha$-helical 
proteins.$^{6,42,43,59}$ More recently, Onuchic et al.$^{9}$ considered
the thermodynamics of a Gaussian random energy model similar to the one 
employed here and derived the relation 
$T_{\rm f}/T_{\rm g}$ $=$ $(\HD/\sH)\sqrt{2/\ln\gcD}$
(in the present notation).\footnote{
Solvent-mediated (effective) intraprotein interactions can have enthalpic
as well as entropic contributions. However, heat-induced conformational
changes would be impossible if these interactions do not contain enthalpic 
parts. The interaction energy $E$ was taken to be purely enthalpic in
Onuchic et al.'s random-energy treatment of temperature dependences
that leads to Eq.~(12) in Ref.~9.} 

The estimate $T_{\rm f}/T_{\rm g}\approx 1.6$ was based on kinetic
simulations. As such, it may be viewed as a lower bound for a protein to 
satisfy a certain requirement for foldability. A previous random-energy-model 
analysis already suggests that a higher thermodynamic
$T_{\rm f}/T_{\rm g}$ ratio may be needed to satisfy the additional
constraint imposed by calorimetric two-state 
cooperativity.$^{23}$ Figure~2b shows calorimetric cooperativity as
a function of $T_{\rm f}/T_{\rm g}$ (the horizonal axis is marked
by the inverse of this ratio, $T_{\rm g}/T_{\rm f}$, by applying Eq.~(12)
of Onuchic et al.$^{9}$). Using realistic protein parameters,$^{23,54}$ 
Figure~2b shows that in the context of the present random-energy model 
analysis, for a protein's $\DHv/\DHc > 0.96$ (Ref.~54), it is 
necessary for $T_{\rm f}/T_{\rm g}>5.8$; $\DHv/\DHc >0.99$ implies
$T_{\rm f}/T_{\rm g}>10.0$; and $T_{\rm f}/T_{\rm g}\approx 1.6$ would imply
that the protein is not calorimetrically two-state, with $\DHv/\DHc < 0.2$.

Therefore, combining our results with Onuchic et al.'s analysis$^{9}$ 
leads us to the conclusion that for proteins that are calorimetrically 
two-state, $T_{\rm f}/T_{\rm g}$ should be higher than the earlier 
estimate of $1.6$, and may well exceed $6.0$.  In that case, even for an 
hypothetical highly stable two-state protein with $T_{\rm f}\approx 100^\circ$C 
(373.15K), $T_{\rm g}$ is still very low, at $\approx 62$K. 
This folding glass transition temperature is a theoretical construct for
quantitating a ``rugged'' landscape's impediment to the kinetics of
folding from the {\it denatured} to the native state. The physics
it describes is different from the ``glass transition'' of {\it native} 
proteins observed experimentally at $\approx 200$K (see, for example, Ref.~61),
though it has been suggested$^{59}$ that the two phenomena might be related.
The present calorimetric estimate of $T_{\rm g}\approx 62$K is 
much lower than temperatures at which folding actually takes place. 
While the idealized enthalpy distribution of a random-energy 
model without explicit chain representation might have underestimated
the chance of having low-enthalpy kinetic traps, 
such traps should nevertheless be improbable given 
this extremely low estimate for $T_{\rm g}$.
Therefore, our results suggest that in general kinetic traps should have 
at most minimal effects on the folding of real calorimetrically 
two-state proteins of sizes comparable to CI2.$^{19,37,42,43}$ 
This view is apparently supported by recent 
folding experiments on proteins with no kinetic intermediates.$^{62-67}$ 
In this perspective, it would be particularly revealing to elucidate
the relationship between multi-phasic kinetics and calorimetric cooperativity
for real proteins that do fold with kinetic intermediates (see, for example, 
Refs.~68--70, and theoretical perspectives in Refs.~3--8, and 11).\\

\noindent
{\bf Lattice protein models: Why compare them against the 
calorimetric two-state criterion?}\\

We now turn to protein models with explicit chain representations.
Recent years have seen sustained efforts in using highly simplified
lattice models to understand general properties of proteins. Lattice
protein models were pioneered by G\=o and coworkers.$^{15}$ G\=o
models assume that only those contact interactions that occur in 
the native conformation can be favorable, whereas all nonnative interactions
are neutral. This approach to modeling may be characterized 
as {\it teleological}, because the native conformation is hardwired explicitly 
into the model potential function. A lot of useful insight has been gained
by this methodology. But it is important to realize that a G\=o model leaves 
open the question as to what physical interactions can conspire to produce 
the remarkable molecular recognition effect it has assumed.

An essential difference between G\=o models and models introduced in the 
past decade --- beginning with the simplest 2-letter HP potential,$^{30,32}$ 
is that many of the more recent models have adopted microscopic 
interaction schemes that are independent of a particular native conformation. 
Therefore, these models offer the possibility to better explore the 
physico-chemical bases of protein folding. While much have been learnt 
(see, for example, Refs.~1--9, 11, 12), the goal of using these models 
to elucidate general protein properties has not been fully realized.
One of the most generic thermodynamic properties of many small single-domain 
proteins is their calorimetric two-state cooperativity. However, no 
three-dimensional lattice model has been evaluated against the 
calorimetric two-state criterion. We do so here for six representative
models. This was motivated by a previous study of two-dimensional 
models,$^{23}$ which has led us to suspect that to design a physically 
plausible three-dimensional interaction scheme to reproduce calorimetric 
two-state behaviors might be non-trivial, and that other deficiencies of 
lattice models in describing real two-state protein properties$^{37}$ might
be intimately related to their lack of calorimetric two-state cooperativity.

We take this as the first step in an endeavor to build simple tractable
self-contained models to capture more proteinlike features.
It is hoped that once models are required to better conform to the 
calorimetric two-state criterion, mechanisms for other two-state proteinlike 
properties would either be apparent or become more easily decipherable. From 
this vantage point, the substantial amount of lattice model data accumulated 
over the years constitutes a valuable repository of information.
By applying appropriate experimental tests 
on these models for their similarities with {\it and} their differences 
from real protein behaviors, one would gain new insight into
what novel energetic ingredients might be necessary for building better 
models.

We consider six models,$^{44-49}$ as shown in Figure~3. We choose to 
analyze these models in depth because they are representative and 
instructive, covering a varieties of approaches and assumptions employed 
in recent efforts to model proteins as chains configured on three-dimensional 
simple cubic lattices. Some models in Figure~3 have been studied extensively 
and contributed significantly to the advances in theoretical understanding. 
All these models are based upon additive pairwise nearest-lattice-neighbor 
contact energies. As described in the original references,$^{44-49}$ 
the contact energies are all assumed to be temperature independent.
We therefore refer to these energies as enthalpies, as in Ref.~23, to
conform to the terminology in the experimental calorimetric literature.\\

\noindent
{\bf Lattice simulation methods.}\\

Using the model potential functions described in their respective
original studies,$^{44-49}$
thermodynamic quantities of these models were computed using standard 
Metropolis Monte Carlo (MC) histogram techniques.$^{71,72}$ The chain move
set we used consists of end, corner, and crankshaft moves, as described
by Socci and Onuchic,$^{44}$ with additional sidechain moves for the 
20-letter sidechain model (Figure~3f).$^{48}$ Each histogram was computed 
using a total of $4.5\times 10^8$ attempted moves, whereby data was 
collected after allowing for an initial equilibrating run of
$5\times 10^7$ attempted moves. Every attempted move is counted as
elapsed MC time in computing Boltzmann averages, whether it is
accepted or rejected; and if rejected, regardless of whether it is caused by
excluded volume violation or by the stochastic Metropolis algorithm for 
an attempted move that involves a finite increase in energy 
(enthalpy). The simulation temperatures are given in 
the captions for Figures~4--9. In one case (the G\=o model in Figure~7), 
we also performed several independent MC simulations at different 
temperatures to confirm the MC histogram results.
Our sampling of the densities of states should be adequate since we 
obtained essentially the same midpoint temperatures as the original studies 
for all six models.\footnote{For the 20-letter model, the temperature at 
which the Boltzmann average $\langle {\bf Q}\rangle$ of the number of 
native contact {\bf Q} equals one half of the total number ${\bf Q}_{\rm N}$ of 
native contacts was reported to be $0.272$ in Ref.~49 (note that this {\bf Q} 
is different from the symbol $Q$ for partition function), whereas the present 
simulation gives $0.279$. The discrepancy is not big. However, it is not clear
whether the discrepancy merely reflects numerical uncertainties or is it 
related to a possible systematic deviation from the correct Boltzmann 
distribution in previous simulations in which attempted moves rejected by 
excluded volume violations were not counted as elapsed MC time (page 185 of 
Ref.~49, page 1617 of Ref.~73), as has been noted recently (Ref.~47).}

Thermodynamic functions relevant to calorimetric considerations are plotted
in Figures~4--9. In these figures, $T_{1/2}$ is the temperature at which the
chain population [N] in the {\it single} lowest-enthalpy conformation
equals $1/2$. This single-lattice-conformation definition of the model native
state and the corresponding identification of $T_{1/2}$ with the folding 
transition temperature coincide with the original formulations in four of 
the models.$^{44-47}$ However, a multiple-lattice-conformation native state 
containing other conformations in addition to the lowest-enthalpy conformation 
was advocated by the authors of the two 20-letter models.$^{48,49}$ 
Hence, according to their definitions, the ``native'' populations in their
models$^{48,49}$ are larger than [N] in Figures~6 and 9. We will give more
detailed consideration to the issue of native state definition below.\\

\noindent
{\bf Evaluating lattice protein models against the calorimetric
two-state criterion.}\\

\noindent
{\it A First Step: Modeling Heat Capacity Functions With No Baseline 
Subtractions}\\

We first apply the model heat capacity and enthalpy functions in 
Figures~4--9 directly to the relation$^{23}$ $\kappa_0$ $=$ 
$\langle\Delta H(T_{1/2})\rangle_{\rm D}/\DHc$ and Eq.~(6) above
to compute various $\DHv/\DHc$ ratios in Table~II. This is equivalent to 
assuming that for each model (as for the random-energy models above), 
the entire model $C_P$ function is directly 
comparable to the ``transition'' part of an experimental excess heat 
capacity function,$^{41}$ the analyses of which has led
to the calorimetric two-state condition $\DHv/\DHc\approx 1$ for many
small proteins. Experimentally, the transition part of the excess 
heat capacity is obtained by performing baseline subtractions on the raw
data.$^{23,41}$ This exercise we now undertake is a necessary and instructive 
starting point that involves minimal assumption,$^{23}$ as it does not
entail performing any baseline subtraction on model results. After a basic 
perspective has been gained, we will discuss in a later section the 
feasibility and appropriateness of applying baseline subtractions to model 
specific heat functions. 

In addition to the $C_P$ functions, the upper panels of Figures~4--9 also 
show the heat capacity contributions $(C_P)_{\rm D}[{\rm D}]$ 
from thermal transitions 
among nonnative (in these cases, non-ground-state) conformations.$^{23}$ 
When a large fraction of $C_P$ arises from transitions among nonnative 
conformations instead of transitions between native (N) and nonnative (D)
conformations, significant deviations from calorimetic two-state 
behaviors by the $\kappa_0\approx 1$ standard are expected$^{23}$ (Table~II).
This is because a large $(C_P)_{\rm D}[{\rm D}]$ contribution means that even
after passing the denaturation transition midpoint (when [D]$>1/2$), the 
average denatured enthalpy will continue to rise substantially when the 
temperature is further raised (see the lower panel of 
Figure~4, for example), as denatured chains are propelled to 
populate conformations at higher and higher enthalpies. Table~II summarizes the
six models' conformity to calorimetric two-state criteria based on 
different $\DHv$'s.  Calorimetric cooperativities measured by common 
experimental $\DHv/\DHc$ formulas such as $(\kappa_2)^2$ and $(\kappa_3)^2$ 
(see Table~I) can readily be calculated from Table~II.\\

\noindent
{\it None of the Models Tested Meets the Calorimetric Two-State Standard}\\

Table~II shows that all six models tested by the present method
do not meet the experimental calorimetric two-state standard. Among
them, the G\=o model appears to be most cooperative, with
$\kappa_0=0.54$ and $\kappa_2$ $\approx$ $\kappa_3$ $=0.87$. If the common
experimental formulas $(\kappa_2)^2$ (Ref.~41) and $(\kappa_3)^2$ (Ref.~51) 
for van't Hoff to calorimetric enthalpy ratio are used, this translates into
$\DHv/\DHc$ $\approx$ $0.75$ for this particular G\=o model. This is still low
when compared with experimental values of $\approx 0.96$ (Ref.~54)  for 
calorimetrically two-state proteins. For five small compact globular
proteins --- ribonuclease A, lysozyme, $\alpha$-chymotrypsin, cytochrome $c$,
and metmyoglobin, Privalov$^{51}$ reported an average $\DHv/\DHc$ $=$ 
$(\kappa_3)^2$ $=$ $0.96\pm 0.03$.\\

\noindent
{\it Different Calorimetric Criteria are Related to Definitions of the
Native State --- 20-Letter Models}\\

For the models tested, the $\DHv/\DHc$ values ($\kappa$'s) vary considerably 
depending on what definition of van't Hoff enthalpy is used (Table~II). 
The variation is mildest for the 2- and 3-letter models, for which
the population-based $\kappa_0$ is almost identical to one of the 
experimental square-root formulas, $\kappa_3$. And while $\kappa_2$'s are 
different from $\kappa_3$'s for these two models, they are only 27--38\% 
larger than $\kappa_0$. For the other four models, the difference between 
$\kappa_0$ and the experimental formulas $\kappa_2$ or $\kappa_3$ is 
larger: $\kappa_3$ is $1.6$ -- $1.8$ times $\kappa_0$ for the G\=o and 
modified HP models, whereas $\kappa_3$ is $\approx 7$ times bigger than 
$\kappa_0$ for the two 20-letter models. For the latter four models, 
however, $\kappa_2$ is virtually identical to $\kappa_3$.

The differences among $\kappa$'s are often related to differences in
the midpoint temperatures used to define them. For the 2- and
3-letter models (Figures~4 and 5), the temperature $T_{1/2}$ for
the population-based $\kappa_0$ (and $\kappa_1$) are well within 
the peak region of the specific heat capacity function and quite 
close to the temperature $T_{\rm max}$ for $\kappa_2$. This accounts 
for the relative small differences among $\kappa_0$, $\kappa_1$, and 
$\kappa_2$ in these models. The difference between $\kappa_0$
and $\kappa_2$ is larger for the G\=o and modified HP models, but 
$T_{1/2}$ still lies within the peak region of the $C_P$ function and
not that far away from $T_{\rm max}$ (Figures~7 and 8). The difference 
between $\kappa_0$ and $\kappa_2$ is much larger for the two 20-letter 
models. In these constructs, $T_{1/2}$ is well outside the peak 
region of $C_P$ ($T_{1/2}\ll T_{\rm max}$, see Figures~6 and 9). On 
the other hand, $T_{\rm max}$ $\approx$ $T_d$ for the G\=o model and 
the 20-letter model without sidechains (Figures~6 and 7), hence 
they have $\kappa_2$ $\approx$ $\kappa_3$.

The large temperature differences between $T_{1/2}$ and $T_{\rm max}$ in 
Figures~6 and 9 highlight one peculiar feature of the two 20-letter models 
which is qualitatively different from the other four models. For both of them, 
the population [N] of the single ground-state conformation is below 10\% at 
$T_{\rm max}$, whereas the $C_P$ at $T_{1/2}$ (when [N] = 1/2) is very low. 
This feature is intimately related to the rationale for adopting a 
multiple-lattice-conformation native state in these models.$^{48,49}$ In 
physical terms, it means that $C_P$ is dominated at low temperatures 
by transitions among the single ground-state 
conformation and other conformations with very low (close to ground-state) 
enthaplies, most of these conformations belong to these models' 
multiple-conformation native state as defined by their authors$^{48,49}$ 
(see below). When the temperature is raised, population in the single 
ground-state conformation continues to decrease as more of it is being 
transferred to other low-enthalpy conformations. Therefore, when the 
temperature reaches $T_{\rm max}$, contributions to the peak value of $C_P$ 
are dominated by transitions between the group of low-enthalpy conformations 
as a whole with the large number of high-enthalpy conformations. By that time 
the population [N] in the {\it single} ground-state conformation has become 
quite insignificant. This is the basic reason why $\kappa_2$ 
$\gg$ $\kappa_0$ $\approx$ $\kappa_1$ for these two models (Table~II).\\

%\vfill\eject

\noindent
{\it Model Heat Capacity Functions can be Compared Directly with 
Experiments --- G\=o and 20-Letter Mainchain Models are More Cooperative}\\

By considering random-energy models, we have argued above that all common 
calorimetric criteria using different $\kappa$'s are essentially equivalent 
when $\DHv/\DHc$ $\approx$ $1$ and the native state is represented by a 
single enthalpy value in an effective density of states that describes
the transition part of an experimental excess heat capacity function 
after proper baseline subtractions.$^{23}$ The behavior of the two 20-letter 
models prompts us to ask a more general question: Which $\kappa$ computed 
from a model would be most relevant for comparing theory with experiment when
$\DHv/\DHc$ deviates significantly from unity and the native state of
the chain model may have multiple enthalpy levels?

From an operational standpoint, among the $\DHv/\DHc$'s considered, 
$\kappa_2$, $\kappa_3$, $(\kappa_2)^2$, and $(\kappa_3)^2$ are most directly 
related to experiments. This is because they can be determined by analyzing 
the model $C_P$ function alone (which corresponds to an experimental 
calorimetric scan) without involving an {\it a priori} definition of the 
``native state'' (whereas such a definition is needed to determine $T_{1/2}$ 
for $\kappa_0$ and $\kappa_1$). It is also prudent to 
not commit prematurely to a general single-lattice-conformation definition 
of the native state.

By this operational standard, the 20-letter model without sidechains is 
second most cooperative after the G\=o model, with $\kappa_2$ $\approx$ 
$\kappa_3$ $=$ $0.66$. On the other hand, the 2-letter, 3-letter and 
modified HP models are far from being calorimetrically two-state by all 
standards considered here: none of their $\kappa$'s exceeds $0.46$; in fact 
they are often much lower (Table~II). In these models, at any one of the 
transition midpoints, the average enthalpic difference 
$\langle \Delta H(T)\rangle_{\rm D}$ between the denatured state and the 
single native conformation is low relative to $\DHc$ (lower panels of 
Figures~4, 5, 9).\\

\noindent
{\it 2- and 3-Letter Models are Less Cooperative --- ``Variable Two-State''
Does Not Equal ``Calorimetrically Two-State''}\\ 

For the 2-letter model in Figures~3a and 4, a previous study 
has shown that its denatured enthalpy distribution is a broad 
shifting peak whose center position is moving continuously to higher values 
as temperature is increased (for example, the peak $H\approx -64$ at $T=1.26$ 
whereas the peak $H\approx -16$ at $T=5.00$, see Fig.~5 of Ref.~72, $H$ is
equivalent to their $E$). Therefore, this 2-letter example corresponds to 
the ``variable two-state'' case of Dill and Shortle (Fig.~1B of Ref.~74) 
with heat (increasing temperature) as the ``denaturing agent.'' The observation 
here implies that the variable two-state scenario can differ substantially 
from a calorimetric two-state transition if it entails significant 
post-denaturational shifting of the enthalpy distribution among the denatured 
conformations. The present calorimetric analysis agrees with previous 
assessments$^{75}$ that the 3-letter model is more cooperative (has larger 
$\kappa$'s, Table~II) than the 2-letter 
model, though both are far from being calorimetrically two-state. We will
consider the 3-letter model in more detail below. The modified HP model
in Figures~3e and 8 was motivated by considerations of hydration 
effects. Its potential function is based on two residue types (H and P), with
novel features$^{47}$ such that it effectively interpolates between 
the standard HP potential$^{30,32}$ (when chain conformations are open) and 
the ``AB'' potential$^{76-78}$ (when chain conformations are compact).
In the AB potential, like residues attract and unlike residues repel.
Repulsive interactions$^{19,77}$ of the AB type facilitate sequence design 
and enhance kinetic foldability in this modified model relative to the 
standard HP model,$^{47}$ though it is insufficient for calorimetric two-state
cooperativity (Table~II). It is interesting to note that the spatial
organization of residues in the native conformation of this modified HP model 
(Figure~3e) is dictated mainly by the AB potential. 
Consequently, the two types of residues are segegrated to opposite
sides of the structure to minimize contact, rather than organizing into
a hydrophobic (H) core surrounded by polar (P) residues as in typical
HP ground-state conformations.$^{79}$\\

\noindent
{\it Short 20-letter Sidechain Models are not Calorimetically Cooperative}\\

We have also calculated Klimov and Thirumalai's$^{48}$ cooperativity parameter
$\Omega_c$ by extending the MC histogram technique to compute the
temperature dependence of their structural overlap function $\chi$ 
(Refs.~23, 48). The results are included in Table~II. While $\Omega_c$ is 
basically a measure of the sharpness of a transition and does not always 
correlate with the 
degree of conformity to calorimetric two-state cooperativity,$^{23}$ for 
these six models the rank ordering of the three most cooperative models 
by $\kappa_2$ coincides with their rank ordering by $\Omega_c$. This 
suggests that $\Omega_c$ may correlate reasonably well with calorimetric 
cooperativity if the conformational entropies of the chain models in 
question are similar. The calorimetric cooperativity as measured by 
$\kappa_2$ and $\kappa_3$ of the 15mer 20-letter sidechain model of 
Klimov and Thirumalai$^{48}$ is low (Figures~3f and 9, their ``sequence A''), 
and is comparable to that of the 2-letter, 3-letter, and the modified HP 
model. Remarkably, by the $\Omega_c$ measure, it is by far the least 
cooperative among the six models.  We have also completed the same analysis 
for their other sidechain model, ``sequence B.'' The results are similar
($\kappa_2=0.25$, other data not shown). The low levels of calorimetric
cooperativity in these sidechain models may be a consequence of the shortness 
of the chains, as it has been observed that models with sidechains on average 
have higher $\Omega_c$'s than non-sidechain models with the same number of 
mainchain monomers.$^{48}$ Nonetheless, the present results mean that how 
sidechains may enhance thermodynamic cooperativity in longer chain models is 
a question that remains to be ascertained.\\

\noindent
{\it The Enthalpy Distribution of G\=o Model is Trimodal}\\

We now take a closer look, as an example, at how the underlying enthalpy 
distribution of the G\=o model (Figures~3d, 7) gives rise to its relatively 
high cooperativity by the calorimetric criterion. Figure~10 shows that
the G\=o model enthalpy distribution is very different from that of
models with much lower cooperativities, such as the 2-letter model of Socci 
and Onuchic.$^{44}$  The enthalpy distribution of the 2-letter model in 
Figures~3a and 4 is bimodal --- the lower mode peaks at the ground-state native 
enthalpy ($-84$) and encompasses enthalpies $<-77$, whereas the higher mode 
has a shifting peak, corresponding to a temperature-dependent variable
enthalpy distribution in the denatured ensemble (Fig.~5 of Ref.~72; see above). 
In contrast, the {\it denatured} enthalpy distribution of the G\=o model 
consists of two widely separated peaks (Figure~10), the lower one is at 
$H=-54$ and the higher one is around $H=-6$ to $-4$. Together with the
native population at $H=-57$, these give rise to a trimodal distribution
of enthalpy. (The native peak is not shown in Figure~10.)

The data in Figures~7 and 10 implies that the heat denaturation of the
G\=o model takes place in the following manner. At low temperatures,
$T<0.5$ for example, $>95\%$ of the chain population is in the single native
conformation (Figure~3d). As temperature is raised to $T=0.65$ -- $0.70$, 
a fraction of the native population is transferred to a group of
low-enthalpy conformations with $H$'s around $-54$ (Figure~10). 
There is an enthalpy (energy) gap of 3 units between the ground state 
and the lowest-enthalpy ($H=-54$) nonnative conformations. Using MC histogram 
techniques, we estimated that there are $\sim 10^5$ nonnative 
conformations with $H<-44$. (For this G\=o model, the number of native 
contacts ${\bf Q}=-H$.) The heat capacity associated with these initial thermal 
transitions is 
small in comparison with the heat absorption peak because of the relatively
narrow enthalpy differences between the native and the low-enthalpy nonnative 
conformations. As temperature continues to increase to $\approx T_{1/2}=0.75$, 
chains start to unfold substantially, and a concentration of population 
at very high enthalpies ($H \approx -6$) begins to develop. This temperature 
coincides with the sharp peak of the heat capacity function (Figure~7, 
upper panel), which reflects the large-enthalpy thermal transitions 
from both the single ground-state conformation ($H=-57$) and the low-enthalpy 
nonnative conformations ($H\approx -54$ -- $-40$) to the large number of 
high enthalpy conformations around $H\approx -6$. There are non-vanishing 
chain populations at enthalpy levels intermediate between the two nonnative 
peaks, but they are not appreciable at any temperature. When the temperature 
is raised further to $T=0.83$ -- $0.95$, the population at the single 
ground-state and the low-enthalpy nonnative conformations greatly 
diminishes and practically all the chains have enthalpies above $H=-16$.\\  

\noindent
{\it Why is the G\=o Model More Cooperative Than Others?}\\

Several features of this process contribute to the G\=o
model's relatively high cooperativity. First, unlike the 
2-letter model discussed above, the population peak of high-enthalpy 
conformations is quite insensitive to temperature: it shifts by merely 
$\approx 2$ enthalpy units, from $H\approx-6$ to $-4$, when the temperature 
is increased from $T=0.75$ to $0.95$ (Figure~10). Second, unlike the 20-letter 
models whose single ground-state conformational populations become $<0.1$ when
the temperature is raised to $T_{\rm max}$ (Figures~6, 9, see above), 
the population of the single-conformation G\=o-model ground state remains
substantial ($\approx 0.3$) at the peak of the heat capacity function.
In fact, all three transition midpoint temperatures are well within the peak 
region of $C_P$ for the G\=o model. And among the models tested, it is 
the one with both $T_{1/2}$ and $T_d$ closest to $T_{\rm max}$ --- 
within 1.4\% and 0.4\%, respectively (Figure~7, upper panel).

These observations rationalize certain differences in cooperativity between 
models. For instance, the G\=o model is more cooperative than the 2-letter 
model in Figure~4 by all $\kappa$ measures in Table~II. This is
because the G\=o model's bimodal distribution of nonnative enthalpies
(i.e., the denatured part of an overall trimodal distribution)
implies that a larger variance in $H$ is possible, hence a higher 
peak value for $C_P$ [Eq.~(2)], and therefore a larger $\kappa_2$,
than the 2-letter model with a single shifting broad distribution of
denatured enthalpies.  The bimodal denatured enthalpy distribution of the 
G\=o model also means that the average denatured enthalpy near
$T_{1/2}$ should be approximately one half of the entire range of 
possible enthalpy variations. Hence $\kappa_0$ should be $\approx 0.5$
(Table~II indeed gives $\kappa_0=0.54$.) This is higher than the $\kappa_0$
of the 2-letter model because the latter's denatured state is dominated 
by low-enthalpy conformations at its $T_{1/2}$. The G\=o model
is more cooperative than the 20-letter model in Figure~6. For the
$\kappa_2$ measure, it is because at $T_{\rm max}$ the G\=o model has 
$\approx 3$ times as much chain population [N] in its single ground-state 
conformation as the 20-letter model. 
The highly specific, teleological interactions of the G\=o model also 
lead to much smaller probabilities for intermediate enthalpies. These 
factors translate into the possibility of having a larger variance in 
enthalpy distribution, thus a higher peak $C_P$ value, and hence a higher 
$\kappa_2$ for the G\=o model than for the 20-letter model.\\

%\vfill\eject

\noindent
{\it Summary of Analysis With No Baseline Subtractions}\\

The analysis above has shown that none of the models tested is 
calorimetrically two-state, though there are wide variations in 
their deviation from being so. For models with relatively high
cooperativities such as the 36mer 20-letter model and the 48mer G\=o model,
this conclusion is still somewhat tentative because baseline subtraction 
schemes$^{22,23}$ are yet to be explored (see below). These schemes can lead 
to effectively higher $\kappa$'s (Ref.~22). However, for models that deviate
far from $\DHv/\DHc\approx 1$ for all van't Hoff enthalpies considered, 
in particular the modified HP and short sidechain models, the analysis carried
out so far is already quite sufficient in establishing that they are not 
good thermodynamic models for real calorimetrically two-state proteins.

It is noteworthy that the present three-dimensional 48mer G\=o model is 
significantly more cooperative by the $\kappa_2$ criterion ($=0.87$) than 
a two-dimensional 18mer G\=o model studied previously ($\kappa_2=0.64$).$^{23}$ 
Apparently, the longer chain length, the ability to form a three-dimensional
core, and even the particular fold topology of the present G\=o model
might have contributed to its higher calorimetric cooperativity. These factors
need to be better elucidated. As we have emphasized, the interactions 
in G\=o models are highly artifical as they are not based explicitly
on a set of plausible microscopic physical interactions. But G\=o model
results are nonetheless instructive as they may highlight intrinsic
limitations to what can be achieved by contact interactions. At least in 
the context of an underlying flexible polymer model, the above observations 
on all six models suggest that there always exists 
conformations with enthalpies (energies) close to the 
ground state, even when conformational distribution is governed by the 
highly specific G\=o potential. This raises the question as to whether it 
is natural to group them together with the ground-state conformation$^{46}$ to 
define a multiple-lattice-conformation native state as advocated by the 
authors of 20-letter models.$^{48,49}$ As will be seen below, 
this is a substantive physical question, not merely an issue 
of semantics. In fact, it
is directly relevant to gaining a better physical understanding of baseline 
subtraction and devising more appropriate means to compare model 
predictions with calorimetric experiments.\\

\noindent
{\bf Effects of discarding a part of model specific heat capacity to mimic
experimental baseline subtractions.}\\ 

\noindent
{\it Physical Meaning of Baseline Subtractions}\\

As a first approximation, we have so far assumed, as in a previous 
study,$^{23}$ that the heat capacity functions predicted by simple protein 
lattice models are directly comparable to the standard ``transition part'' of 
experimental excess heat capacity function. The latter were obtained from 
calorimetric data by subtracting a sigmoidal weighted baseline after
first subtracting the buffer baseline.$^{23,36,41}$ 
This follows from the conventional experimental interpretation$^{33,36,51}$
that only the peak region of $C_P$ involves appreciable heat capacity 
contributions from thermal transitions between conformations that are both 
structurally and enthalpically significantly different from one another. 
In this conventional view, by subtracting the baselines, the heat capacity 
contributions discarded were essentially only those from solvation effects 
and small-amplitude motions of the protein, i.e., contributions that are
regarded as unimportant in accounting for significant conformational changes.
This assumption also underlies the standard empirical approach of using 
temperature-independent solvent accessible surface areas for both the folded 
and the unfolded states of a protein in thermodynamic analyses of 
calorimetric data.$^{33,36}$ However, this picture does not correspond
exactly to the properties of polymer protein models, which invariably
predict a non-negligible heat capacity contribution from conformational
transitions well above the peak $C_P$ transition region, though the 
amount of this contribution varies from model to model (see below).

There are other reasons to believe that the real physical situation
may be more complex than the picture implied by our first approximation and
conventional empirical interpretation of calorimetric data.
Bond vector motions measured by NMR spin relaxation indicate that protein
backbone fluctuations contribute 8 -- 14 cal mol$^{-1}$K$^{-1}$ per 
residue,$^{80,81}$ and thus account for $\sim 20\%$ of the heat capacity 
of an unfolded protein. On the other hand, similar measurements on
the folded state of two proteins suggest that backbone fluctuations on
average contribute only 0.5 cal mol$^{-1}$K$^{-1}$ per residue, and account 
for $\sim 1\%$ of the heat capacity of the native state. While the
connection between NMR-measured bond vector motions and conformational
diversity remains to be better elucidated, the huge difference in
heat capacity contribution from backbone motions between the folded
and unfolded states strongly suggests that the possibility of enthalpic 
transitions between structurally dissimilar conformations in the denatured 
ensemble cannot be neglected, and that conventional baseline subtractions 
might have discarded heat capacity contributions from these transitions.

More recently, a molecular dynamics simulation study using implicit solvent 
interactions also suggests that in addition to differences in solvation 
effects, there are significant heat capacity contributions to the difference 
between native and denatured baselines from noncovalent intraprotein 
interactions.$^{38}$ While the heat capacity contributions from model 
vibrational motions of the covalent bonds$^{82}$ are essentially the same 
in the native and the denatured states, these simulations suggest that 
noncovalent interactions change more with temperature in nonnative 
conformations than in the native state.$^{38}$ Owing to limited sampling,
large numerical uncertainties were reported in this molecular dynamics
study. Nonetheless, its prediction that on average non-solvation 
intraprotein interactions account for $\sim 71\%$ of the heat capacity 
difference between native and denatured baselines (Table~2 of Ref.~38) 
appears to be consistent with the NMR experiments described above: If we 
perform a rough estimate based on cytochrome $c$ data (Ref.~33), and
take $\sim 16$ -- $23$ cal mol$^{-1}$K$^{-1}$ per residue to be typical 
native-denatured baseline differences, the NMR results$^{81}$ suggest
that $\sim 50$ -- $70\%$ of this difference may originate from the difference
in backbone motions in the native vs. the denatured state, which is
in the same range as the average molecular dynamics result.

From a polymer perspective, it is also intuitive to expect 
non-vanishing heat capacity contributions from thermal transitions
between conformations at different enthalpic levels with 
significant structural differences even at temperatures above the
peak $C_P$ region. Given the immense diversity in conformational 
structures, it is physically quite inconceivable how enthalpic diversity 
in the denatured ensemble can be entirely eliminated such that it behaves 
as if all conformations occupy only a single enthalpy level, which would
have meant that all intraprotein solvent-mediated interactions in the 
denatured ensemble were exclusively entropic.

Among the lattice protein models evaluated here, in which we have taken
all interactions to be enthalpic for simplicity, even the heat capacity
function of the G\=o model with relatively high calorimetric cooperativity 
has a long high-temperature tail (Figure~7, upper panel). This indicates that 
for this model, non-vanishing contributions to $C_P$ from conformational 
transitions are not negligible at high temperatures. A relatively 
long (native) tail extending to temperatures far lower than the peak
$C_P$ region is also present for the two 20-letter models (Figures~6, 9, 
upper panels). On the other hand,
in conventional analyses of calorimetric data, no such long tails are ever 
present to be considered in the transition part of the excess heat capacity 
function obtained from baseline subtractions.$^{36,41,50,51,54}$ Even in
calorimetric analyses of non-cooperative nonprotein homopolymers,$^{83}$
their existence is routinely precluded by empirical baseline 
subtraction techniques. This mismatch between theoretical predictions and 
standard transition excess heat capacities necessitates a closer 
examination of the correspondence between the physical pictures emerging 
from polymer protein models and the conventional interpretation of 
calorimetric experiments.\\

\noindent
{\it Applying Baseline Subtractions to Model Heat Capacities Can Result
in Higher Predicted Calorimetric Cooperativities}\\

We now explore the effects of using an {\it ad hoc} empirical procedure, 
similar to what has been carried out on experimental calorimetric data, to
eliminate both the native and denatured tails in model $C_P$ functions
plotted in the upper panels of Figures~4--9.
Physically, this exercise was motivated by our recognition, based on
the evidence above, that conventional calorimetric baseline analyses 
might have substracted out ``tail'' contributions that are relevant for
the evaluation of polymer model predictions. Hence, as an effort to 
put theoretical predictions on the same footing as the (no-tail) experimental 
transition excess heat capacities, we now perform baseline subtractions on 
model data to eliminate their tail contributions. We do expect, nonetheless, 
that the corresponding ``tail'' contributions in real experimental data are 
only a minor part of the heat capacity contributions discarded by conventional 
baseline subtraction on calorimetric measurements. There are reasons 
to expect that conventional interpretation
is at least partially correct in that a majority of the contributions 
subtracted by standard baseline analyses are indeed heat capacity 
contributions from solvation effects and small-amplitude protein motions.
For instance, the molecular dynamics simulation discussed above 
estimated$^{38}$ that only $\sim 11\%$ of native-state heat capacity came
from non-covalent interactions.

Following standard experimental procedures,$^{50,51}$ (see also Ref.~22)
baselines are constructed as plausible linear extrapolations from
low temperature and high temperature parts of the $C_P$ function 
to its peak region; they are referred to as native (low temperature) and 
denatured (high temperature) baselines. These constructions are depicted
in Figure~11 and the upper panels of Figures~12 and 13 for the six lattice
protein models we have been considering. More details are described in 
the caption for Figure~11. Baseline subtraction has two opposite effects on the 
predicted calorimetric cooperativity. On one hand, it decreases the value 
of calorimetric enthalpy, because some areas under the $C_P$ curve are 
excluded from the integration for $\DHc$. This tends to increase the 
$\DHv/\DHc$ ratio. On the other hand, it decreases the effective peak value 
of heat capacity. This tends to decrease the $\DHv/\DHc$ ratio, as $\DHv$ is 
proportional to the effective peak $C_P$ value or its square root. Here 
we define an effective post-baseline-subtraction $\DHv/\DHc$ ratio by
substituting the new effective peak heat capacity and effective calorimetric
enthalpy into the expression for $\kappa_2$ in Eq.~(6):
$$
\kappa_2\rightarrow \kappa_2^{\rm (s)}\equiv
{\frac 
{2T_{\rm max}\sqrt{\kB C_{P,{\rm max}}^{\rm (s)}}}
{{\DHc}^{\rm (s)}}} \; . 
\eqno(9)
$$
Table~III shows that for all six models, baseline subtractions lead
to increases in apparent (effective) calorimetric cooperativity. 
However, both the modified HP model ($\kappa_2^{\rm (s)}=0.41$)
and the short 20-letter sidechain model ($\kappa_2^{\rm (s)}=0.54$)
remain very far away from being calorimetrically two-state, despite
some improvements. On the other hand,
the effective calorimetric cooperativities of the 2- and 3-letter models
increase dramatically (from $\kappa_2=0.36$ and $0.46$ to 
$\kappa_2^{\rm (s)}\approx 0.94$) after large areas (thick denatured tails) 
under their $C_P$ functions have been subtracted out 
(Figure~11a and upper panel of 
Figure~12). Remarkably, the G\=o model's $\kappa_2^{\rm (s)}$ of $1.00$ 
now meets the experimental standard. The 36mer 20-letter 
model's $\kappa_2^{\rm (s)}$ also rises above $0.94$ (upper panel of 
Figure~13). We will use the 27mer 3-letter and the 36mer 20-letter models 
to discuss the physical implications of these enhancements 
of apparent calorimetric cooperativity by baseline subtractions.\\

\noindent
{\it Nonlinear ``Formal Two-State'' Baselines and Multiple-Conformation
Native States}\\

Recently, Zhou et al. made a pertinent observation that any density of 
states can be formally decomposed into two arbitrary ``states,''
and that its thermal behavior made to satisfy the calorimetric two-state
criterion if one is willing to introduce (non-standard) nonlinear 
baseline subtractions.$^{22}$ To gain further insight into the physical
meaning of baseline subtractions, we found it instructive to contrast 
and compare the present {\it empirical} analysis with their construction.
Here is a brief summary of their formulation (in our notation).
Any partition function $Q$ can be written as a sum of a pair of partition 
functions for two ``states,'' denoted here
by ``0'' and ``1''; viz., $Q(T)=Q_0(T)+Q_1(T)$. Let $(C_P)_0$ and $(C_P)_1$ be 
the individual heat capacities of the two states, computed from $Q_0$ and 
$Q_1$ respectively, and $T_m$ be the midpoint temperature at which the 
population in the two states are equal, i.e., $Q_0(T_m)=Q_1(T_m)$.
Zhou et al.'s baselines are defined by the individual heat capacities:
$(C_P)_0(T)$ for $T<T_m$ and $(C_P)_1(T)$ for $T>T_m$. Naturally,
a calorimetric enthalpy $\Delta^1_0 H_{\rm cal}$ is defined to be
the area between the $C_P$ curve and this baseline, and a midpoint
heat capacity value 
$\Delta^1_0 C_P(T_m)\equiv$ $C_P(T_m)$ $-$ $[(C_P)_0(T_m)+(C_P)_1(T_m)]/2$.
A population-based van't Hoff enthalpy $\Delta^1_0 H_{\rm vH}(T)$ is then 
computed using Eq.~(3) above with $\theta=$ $Q_1(T)/Q(T)$.
Zhou et al. showed that in general 
$\Delta^1_0 H_{\rm vH}(T_m)$ $=$ $\Delta^1_0 H_{\rm cal}$
$=$ $4\kB T^2_m \Delta^1_0 C_P(T_m)/\Delta^1_0 H_{\rm cal}$
[Eqs.~(3), (4), (12) and (15) of Ref.~22]. This identity, which corresponds 
to $\kappa_0$ $=$ $(\kappa_1)^2=1$ if $T_{1/2}$ is formally replaced 
by $T_m$ [see Eq.~(6)], means that the calorimetric two-state 
condition is always satisfied with this particular choice of baselines.

We have computed baselines for the six models according to this
recipe$^{22}$ and included them as dotted curves in 
Figure~11 and the upper panels of Figures~12 and 13. (In the discussion 
below, they are referred to simply as ``nonlinear baselines.'')
For models that assume a single-conformation native state,$^{44-47}$
$Q_0=Q_{\rm N}$ and $Q_1=Q_{\rm D}$. For the two 20-letter models, 
$Q_0$ is constructed as the partition function for the multiple-conformation 
native state defined by the original authors,$^{48,49}$ while $Q_1$
is defined to account for the rest of the conformations. These nonlinear 
``formal two-state'' baselines are conceptually enlightening (see below),
however, it is our view that they should not be used directly to evaluate 
protein models. The first reason is logical --- since 
by construction they always lead to perfect agreements with the calorimetric 
two-state condition, using them on model systems would abolish the 
substantive physical question of whether polymer protein models conform to the 
experimental calorimetric requirements. Second, and more importantly, such
baselines had not been used by experimentalists to analyze calorimetric
data. For all cases studied here, these nonlinear baselines invariably 
subtract more from the peak $C_P$ region than conventional linear or weighted
baselines (Figures~11--13). This means that using these nonlinear 
baselines on model $C_P$ functions would most likely lead to an effective heat
capacity function that does not physically match the experimental 
transition excess heat capacity function,$^{41}$ and thus would make
it extremely difficult to conduct meaningful comparisons between 
theory and experiment.$^{23}$

Much insight can be gained, however, by comparing the nonlinear baselines
with the {\it ad hoc} empirical linear baselines we used. As the nonlinear
baselines of Zhou et al.$^{22}$ are guaranteed to produce perfect (apparent)
calorimetrically two-state behaviors, it is not unreasonable to expect that 
if the linear baselines are close to the nonlinear baselines, the
apparent calorimetric cooperativity predicted by the linear baselines
would be high, and vice versa. This appears to hold for five out of
our six cases: Relatively high apparent calorimetric cooperativities
resulted from linear baseline subtractions for the 2-letter, 3-letter, 
and 36mer 20-letter models (Table~III); and as expected their linear and 
nonlinear baselines are quite close (Figure~11a, upper panels of 
Figures~12 and 13). On the other hand, the nonlinear baselines are
very far away from the empirical linear baselines used for the
modified HP and the 15mer sidechain models. Not surprisingly, their apparent 
calorimetric cooperativities remain low even after linear baseline subtractions 
(Figures~11c and d and Table~III). 

The only exception is the G\=o model (Figure~11b), for which the nonlinear 
denatured baseline amounts to a dominant contribution to the overall heat 
capacity, and is very far from the empirical linear denatured baseline. Yet 
the G\=o model is the most cooperative among the models we evaluated, 
especially after linear baseline subtractions (Table~III). The reason for 
this behavior is because we have taken the denatured state of this model 
to be the ensemble that encompasses all non-ground-state 
conformations. And since the enthalpy distribution among these nonnative 
conformations is bimodal (Figure~10), the nonlinear denatured baseline, 
which is the denatured heat capacity $(C_P)_1=(C_P)_{\rm D}$, involves large 
thermal transitions between the two denatured peaks. This accounts for its 
high magnitudes. In addition, owing to the adoption of a single-conformation 
native state, there is no nonlinear native baseline in the present 
consideration of this G\=o model.  On the other hand, {\it if} 
a multiple-conformation native state were adopted to incorporate the 
low-enthalpy conformations that are now 
being classified as denatured, it would have resulted 
in nonlinear baselines for {\it both} the alternately defined native and 
denatured states. Adoption of such a multiple-conformation native state 
would lead to the elimination of contributions to $(C_P)_1$ from large 
thermal transitions between the two enthalpy peaks in Figure~10, and hence 
a nonlinear denatured baseline with much reduced magnitudes. It is expected
that the nonlinear baselines would then be much closer to the empirical
linear baselines used in our analysis, and would give rise to a situation much
more similar to the 36mer 20-letter case, to be discussed below.

For the 36mer 20-letter model (Figure~13), the (low temperature) nonlinear 
native baseline derived from a multiple-conformation definition$^{49}$ of the 
native state is almost identical to the empirical linear native baseline.
By construction, a nonlinear native baseline accounts for the heat capacity 
contribution from thermal transitions among the multiple conformations of 
the native state. Therefore, when an empirical
linear native baseline essentially overlaps a particular nonlinear native
baseline, and we use the empirical linear baseline for subtraction, we are
effectively (empirically) adopting the multiple-conformation native state 
that underlies the construction of the given nonlinear native baseline.
More generally, when empirical linear baselines for {\it both} the native and
denatured states overlap significantly with their nonlinear counterparts
for a particular formal two-state definition,$^{22,49}$ and $T_{\rm max}$ 
$\approx$ $T_m$, as in this particular 20-letter case (Figure~13, upper 
panel), the empirical linear baseline subtraction scheme may be viewed
as an empirical (approximate) adoption of the given 
formal two-state definition for the native and denatured states.
Hence, it follows from the ``formal two-state'' consideration$^{22}$
that such an empirical subtraction would
lead to closer conformity to the calorimetric two-state criterion
as observed here.\\

\noindent
{\it The 3-Letter (3LC) Model Predicts Significant Post-Denaturational 
Chain Expansion --- Comparison with SAXS Experiments}\\

We now broaden our attention to other thermodynamic properties. Obviously,
adherence to the calorimetric two-state criterion is only one of 
many physical properties of real two-state proteins. Therefore, to ascertain
whether a model with high apparent calorimetric cooperativity is adequate 
for generic properties of real two-state proteins, we should also
subject its other properties to 
further experimental evaluation. In this spirit, we now consider the 
3-letter model in more detail. This model uses a single-conformation
native state,$^{45}$ and its apparent calorimetric cooperativity is quite high
after empirical baseline subtractions, $\kappa_2^{\rm (s)}=$ $0.952$. 
Its behavior is expected to be representative of lattice protein models 
that are based on additive pairwise contact energies and have small numbers 
of monomer (residue) types in their alphabets. For instance, 
in many respects the 
properties of the 3-letter model are similar to the 
2-letter model, which also attains a high apparent calorimetric 
cooperativity after baseline subtractions (Table~III). As discussed above,
the 3-letter model is instrumental in Onuchic et al.'s
$T_{\rm f}/T_{\rm g}$ $=$ $1.6$ estimate for small $\alpha$-helical
proteins.$^{59}$

One thermodynamic property accessible to experimental determination
is the dimension of a protein, measured by its average (i.e., mean-square) 
radius of gyration $R_g$ as a function of temperature. Using the MC histogram 
method, we have computed this function for the 3-letter model 
(Figure~12, lower panel). It shows a very gradual 
post-denaturational increase (for $T>T_{1/2},T_{\rm max}\approx 1.5$):
Average $R_g$ is $\approx 30\%$ larger at higher temperatures than its 
value at the high-temperature edge ($T\approx 1.8$) of the peak $C_P$ 
transition region.

It appears that this prediction is signficantly different from 
experimental observations. Sosnick and Trewhella$^{84}$ have used 
small-angle X-ray scattering (SAXS)
to monitor the temperature dependence of $R_g$ of ribonuclease A,
one of the first few proteins shown to be calorimetrically
two-state.$^{50}$ They observed no systematic post-denaturational
increase of $R_g$ under both reducing (no disulfide bonds) and non-reducing 
conditions. Under reducing conditions (which more closely corresponds
to the present lattice chains without crosslinks), the transition 
temperature $\approx 51^\circ$C. Sosnick and Trewhella observed 
no continuous chain expansion at temperatures higher than the relatively 
narrow transition region at $\sim 45$ -- 54$^\circ$C. Indeed, there was 
even a slight decrease in $R_g$ when the temperature reached 74$^\circ$C.
More recently, Hagihara et al.$^{85}$ used solution X-ray scattering
to show that the temperature dependence of $R_g$ during heat 
denaturation of ribonuclease A and cytochrome $c$ can be well approximated
by a strictly two-state model. Plaxco et al.$^{64}$ used SAXS to monitor the 
dependence of $R_g$ of protein L on guanidine hydropchloride concentration. 
They also did not observe any trend of post-denaturational expansion.

The significant post-denaturational chain expansion predicted by
the 3-letter model is directly related to a substantial heat-induced
shifting of its denatured enthalpy distribution, as evident from
its thick high-temperature $C_P$ tail. This behavior
is similar to that noted above for the 2-letter model.
The discrepancy between this 3-letter model's $R_g$ prediction and 
experiment\footnote
{
Our conclusion here is based on the fact that the 3-letter model $R_g$ 
continues to increase as the temperature is raised above the peak $C_P$ 
transition region, and that this behavior is not observed in experiments. 
Following this logic, if the subtraction scheme in Figure~12a is used to ensure
high calorimetric cooperativity, there should be no appreciable increase 
in model $R_g$ for $T>2.2$ if the prediction is to be consistent with 
experiment.  But this is not the case (Figure~12b). We believe
this reflects the main physical difference between this model and
experimental observation. We note, however, that a direct mapping
of temperatures between the 3-letter model results and experiment
is not possible because they are systems of very different sizes.
For instance, the peak $C_P$ transition regions for real proteins
cover a range of 10 -- 20 degrees (Refs.~50, 84). However, if we choose
an energy unit to equate the 3-letter model $T_{\rm max}\approx 1.51$ 
with the ribonuclease A midpoint temperature of 51$^\circ$C,
the model peak $C_P$ transition region would translate into a temperature
range of $\approx 130$ degrees.
}
suggests that, in spite of its relative high apparent 
calorimetric cooperativity after empirical baseline subtractions,
it suffers from essential deficiencies as a model for real 
two-state proteins because of the broad and shifting enthalpy
distribution among its denatured conformations.\footnote
{
Incorporation of empirical baseline subtractions does not change
our previous conclusion that additive hydrophobic interactions
are insufficient for calorimetric two-state cooperativity.$^{23}$
For the two-dimensional HP, G\=o and HP+ models analyzed in Ref.~23,
application of empirical baseline subtractions similar to the one
used here is not sufficient for bringing their apparent van't Hoff to
calorimetric enthalpy ratio close to unity. However, baseline subtractions 
are able to take the two new models introduced in Ref.~23 with cooperative
interactions much closer to apparent calorimetric two-state behaviors: 
After subtraction, $\kappa_2^{\rm (s)}=$ $0.90$ for the new cooperative
model with pure 
enthalpic interactions, and $\kappa_2^{\rm (s)}=$ $0.97$ for the model 
with entropic HH interaction in Ref.~23. The present consideration of 
the 3-letter model also generalizes the previous observation that HP-like 
nonspecific pairwise additive interactions are insufficient to account 
for certain generic thermodynamic properties of real two-state proteins.
}
This observation is consistent with the proposal above that the ratio
$T_{\rm f}/T_{\rm g}$ $=$ $1.6$ deduced from the 3-letter model
may most likely be an underestimate for real two-state proteins.\\

\noindent
{\it Multiple-Conformation Native State and Non-Native Contacts in the
20-Letter Model}\\ 

Finally, we examine in more detail the thermodynamic of 
the 36mer 20-letter model (Figures~13--15). This model has an apparent 
calorimetric cooperativity ($\kappa_2^{\rm (s)}=0.943$) 
similar to that of the 2- and 3-letter models (Table~III).
Its model potential is the basis of a large body of 
interesting work;$^{11}$ and is expected to be representative of lattice 
protein models that are based on additive pairwise contact energies,
with a large but finite number of monomer types in its alphabet, and a 
substantial fraction of the contact interactions being repulsive.$^{19,72,77}$ 
Here it also serves to exemplify models with a multiple-lattice-conformation 
native state.\footnote{ 
Note, however, that a single-conformation native state with a single
ground-state energy $E_{\rm N}$ was used by the author of Ref.~11 to define 
the folding transition temperature $T_{\rm f}$ for a different lattice 
model in Ref.~76.}

The lower panel of Figure~13 shows how the folding/denaturation transition
of this model chain is tracked by different thermodynamic order parameters,
which may correspond to different experimental probes. The population [N] of
the single ground-state conformation begins to drop rapidly well below
the $C_P$ peak temperature $T_{\rm max}$, whereas $T_{\rm max}$ essentially 
coincides with 
the midpoint temperatures for all other probes shown. This is consistent with 
the observation$^{19}$ that in general the midpoint temperature for [N] is 
lower than that for $\langle{\bf Q}\rangle$. The measure $P({\bf Q}>20)$
shows the sharpest transition, as it is a binary ``formal two-state''
order parameter for which a chain conformation can take only one of two values:
either it is native (has ${\bf Q}>20$), or not (${\bf Q}\le 20$).\footnote{
Using MC histogram technique, we estimated that there are 
$\sim 4.4\times 10^9$ different conformations in this 20-letter sequence's 
${\bf Q}>20$ native state. This is $>10^4$ times more than the $\approx 10^5$
low-enthalpy conformations in the G\=o model (see above), notwithstanding 
a 48mer G\=o model's total number of conformations is 
$\approx (4.68)^{(48-36)}=1.1\times 10^8$ times that
of this 36mer model.$^{86}$ This shows that 
if a multiple-conformation native state were to be defined for the
G\=o model, its conformational diversity would be much smaller than
the one in this 20-letter model.}
The order parameter
$\langle{\bf Q}\rangle/{\bf Q}_{\rm N}$ shows a broader transition because 
there are 40 possible {\bf Q} values for this 36mer chain. For this model,
the temperature dependence of $\langle\chi\rangle$ correlates almost perfectly 
with that of $\langle{\bf Q}\rangle$ (see inset in the upper panel
of Figure~13). These observations illustrate that the sharpness of a 
transition$^{48}$ can vary significantly depending on the probe (order 
parameter), whereas the calorimetric criterion is a more fundamental measure
of cooperativity$^{33}$ because it directly probes the underlying density of 
enthalpic states.$^{23}$

This 20-letter model is a better mimic of real two-state proteins than 
the 3-letter model in certain respects. For instance, its $R_g$ shows no 
significant post-denaturational expansion and therefore enjoys better 
agreement with the SAXS experiments discussed above (Figure~15, lower panel).
We now briefly touch on two issues that are likely to be relevant in future 
assessments of the 20-letter model's conformity to experimental two-state 
behavior.  (i) Structural diversity of the native state: The 20-letter
model allows for significant conformational variation (Figure~14). 
For this particular sequence, this leads to the prediction that the native 
state has a higher heat capacity contribution from main-chain-like motions than
the fully unfolded state, as is evident from the higher $C_P$ value in the 
native tail region than the denatured tail region (Figure~13, upper 
panel).\footnote{A recent G\=o-like continuum three-helix bundle model 
also predicts a higher heat capacity for the native state than the 
denatured state.$^{22}$}
However, this does not appear to agree with the NMR experiments discussed
above.$^{81}$ (ii) The prevalence of nonnative contacts: For this
model, the number of nonnative contacts undergoes a sharp transition
near the heat absorption peak (Figure~15, upper panel). The average number 
is $>3$ at $T_{\rm max}$, reaches a peak $\approx 6$ at a temperature
slightly higher than $T_{\rm max}$, then settles down gradually at a
relatively high average number of $\approx 4.5$ for the high-temperature
unfolded state. Recent NMR experiments show that nonnative interactions 
can exist in the compact denatured states of some proteins,$^{87,88}$ but
this phenomenon is not universal.$^{89}$ If prevalence of nonnative
contact is not a generic property of denatured states of real two-state 
proteins, it would be important to ascertain whether the high number
of nonnative contacts observed in this particular sequence reflects a general 
feature of its underlying 20-letter contact potential.\\ 

%\vfill\eject

\centerline{\large\bf Concluding Remarks}

$\null$

We have examined the implications of calorimetric two-state cooperativity 
and other experimentally determined thermodynamic properties
on a protein's density of enthalpic states.$^{23,90}$ 
In general, they require a narrow enthalpy distribution among the denatured 
conformations, as has been recently proposed.$^{23}$ Energy landscape 
theory$^{9}$ has allowed us to make a connection between calorimetric 
two-state cooperativity and folding kinetics. Using an
analytical random-energy energy model, we showed that the folding
landscape parameter $T_{\rm f}/T_{\rm g}\approx 6.0$, which is significantly 
higher than a previous estimate of $\approx 1.6$ for small 
($\sim 60$-residue) $\alpha$-helical proteins.$^{59}$ Experimental
observations of single-exponential folding without kinetic trapping
for a number of small single-domain proteins 50--80 residues 
long with no disulfide bonds$^{62-67,91-93}$ is consistent with either 
$T_{\rm f}/T_{\rm g}\approx 1.6$ or $\approx 6.0$. This is because for 
proteins with $T_{\rm f}<100^\circ$C, both ratios imply a $T_{\rm g}$
far lower than any temperature at which folding kinetic experiments have
been conducted ($T_{\rm g}<233$K or $<62$K). In general, the present
random-energy-model results also imply that folding of {\it all} 
calorimetric two-state proteins should not be affected by kinetic traps. 
However, this does not appear to agree with experiment. 
Notable counter-examples 
include the calorimetrically two-state$^{33}$ lysozyme$^{94,95}$ and 
cytochrome $c$.$^{96}$ This underscores an intrinsic limitation of the 
random-energy-model method because it is not a chain-based approach and does 
not address sequence-specific properties.

We have evaluated six lattice protein models against the calorimetric
two-state criterion. The initial stage of our analysis treated the native 
state as a single lattice conformation. This was based on the assumption 
in conventional analyses of calorimetric data, which have identified 
the native state as the structure deposited in the Protein Data 
Bank.$^{33,36}$ Therefore, as in a previous investigation,$^{23}$ we 
first evaluated $\DHv/\DHc$ ratios directly from the model $C_P$ functions, 
without any baseline subtractions (i.e., the baseline was first taken to 
be simply the $C_P=0$ axis).
In this evaluation, none of the models came close to meeting the calorimetric 
two-state standard. This is consistent with our previous conclusion,
based on two-dimensional models, that when the native state is considered
to be consisting of a single conformation, pairwise additive contact 
interactions are insufficient for calorimetric two-state cooperativity.$^{23}$ 

However, based on both theoretical and experimental considerations, 
principally data from NMR bond vector motion measurements,$^{81}$
we have come to believe that it would be profitable to explore using 
empirical linear (nonzero) baselines to subtract out ``tail contributions'' 
from model $C_P$ functions so as to compare them on a more equal footing with 
experimental transition excess heat capacity functions. We have therefore
taken the second step of incorporating empirical baseline subtractions in
our model evaluation. Analysis
of a 20-letter lattice model indicates that subtracting a nonzero native 
baseline amounts to a re-definition of the native state. Physically, the 
empirical subtraction operation is roughly equivalent to (i) classifying more 
conformations as native, (ii) including their contributions in the 
thermodynamic properties of a multiple-conformation native state, 
and (iii) excluding thermal transitions among these multiple native 
conformations from contributing to the subtracted heat capacity function. 

After baseline subtractions, a G\=o model 
meets the calorimetric two-state standard. However, while the 
teleological G\=o potential is extremely useful for posting 
``what if'' questions,$^{43,46}$ whether and how it can be rationalized 
in terms of physically plausible interactions remains to be clarified.  
Among models with a finite alphabet of residue types, the apparent 
$\DHv/\DHc$ ratio for the 36mer 20-letter model is relatively high
after empirical baseline subtraction, though it still falls 
short of meeting the high experimental standard for two-state cooperativity.
(Its $(\kappa^{({\rm s})})^2=0.89$, the corresponding 
ratio for real two-state proteins $\approx 0.96$.)
Other models with smaller alphabets or shorter chain lengths
either have low $\DHv/\DHc$ ratios or exhibit significant 
post-denaturational chain expansions that appear to contradict 
X-ray scattering experiments.$^{84,85}$ This suggests that a relative high
level of interaction heterogeneity --- as characterized by a larger
alphabet$^{11,97-99}$ and the presence of repulsive interactions$^{19,72,77}$ 
--- is necessary for more proteinlike thermodynamic cooperativity.

The low-temperature tails in the $C_P$ functions of the 36mer 20-letter
and the G\=o models before baseline subtractions are direct consequences
of the low-enthalpy conformational diversity embodied in the 
multiple-conformation native state of the 36mer 20-letter model, and 
the existence in the G\=o model of $\sim 10^5$ conformations with enthalpies 
very close to its ground
state. This suggests that, for flexible heteropolymer models that achieve high 
apparent calorimetric cooperativity with only pairwise additive contact 
interactions, the native state effectively defined by an empirical 
native baseline would inevitably involve significant conformational 
fluctuation (as modeled here by different discrete lattice conformations).
If one assumes that this model prediction captures at least partially the 
properties of real proteins, this would imply that the {\it a posteriori}
experimental calorimetric ``native state'' defined operationally by 
empirical baseline subtractions may involve significant conformational 
diversity, and therefore may be qualitatively different from the 
{\it a priori} single-conformation native state used in conventional 
interpretation.$^{33,36}$

One of the main goals of this study was to ascertain the degree to
which proteinlike thermodynamic cooperativity can be achieved by
simple models, especially the question as to whether pairwise
additive contact interactions are sufficient. This is part of an effort
to delineate the extent to which existing simple protein models capture
generic protein properties.$^{37}$ This issue is also relevant
to a related question regarding the sufficiency of contact interactions
for protein structure prediction.$^{100}$ Our analysis of the 36mer
20-letter model is particularly instructive. Its apparent calorimetric 
cooperativity is relatively high after empirical baseline subtractions.
However, how well does its predicted native conformational diversity 
match that in real proteins remains to be further investigated, especially
in view of the apparent discrepancy between NMR main-chain bond vector 
motion measurements and the relative magnitudes of the native and unfolded 
heat capacities in this model.

Conventional interpretation of calorimetric data has been premised on a 
single-conformation, X-ray crystal-structure-like native state. The present 
analysis suggests a new perspective that involves a higher degree of 
conformational heterogeneity, namely (i) the possibility of a 
multiple-conformation native state, and (ii) the possibility that conventional 
baseline subtractions could have masked a non-negligible post-denaturational 
change in chain dimension driven by thermal transitions among denatured
conformations
at different enthalpic levels. In this alternate scenario, the relationship 
between calorimetric two-state cooperativity and a protein's underlying 
enthalpic density of states becomes more complex. Nonetheless, 
if one characterizes the thermodynamics of real two-state proteins by
both the calorimetric two-state criterion and the experimental 
observation$^{84,85}$ that no significant post-denaturational 
chain expansion took place, one central aspect of the physical picture$^{23}$
remains essentially the same: For thermodynamically two-state proteins,
there is no significant post-denaturational shifting of the enthalpy 
distribution among the conformations of the denatured state relative to 
the average enthalpy of the (multiple-conformation) native state.
On the other hand, a corresponding {\it pre}-denaturational shifting 
(i.e., under native conditions) does not contradict the experimental 
observations. This is consistent with the multiple-state picture$^{101,102}$
emerging from native-state hydrogen exchange experiments,$^{103,104}$
as has been discussed.$^{23}$ However, it is noteworthy that the baseline 
analysis in the present work does raise the possibility 
that parts of the structural fluctuation revealed by native-state hydrogen 
exchange can in principle correspond to conformational diversities that 
have been operationally absorbed into the baseline-defined 
calorimetric native state.

%*******************************************************************
$\null$\\

\noindent
{\bf Acknowledgments}\\
We thank Yawen Bai, Wayne Bolen, Julie Forman-Kay, Ernesto Freire, 
Roxana Georgescu, Lewis Kay, Ed Lattman, Themis Lazaridis, Kip Murphy, 
Kevin Plaxco, Nick Socci, Tobin Sosnick, Mar{\'i}a-Luisa Tasayco,
and Dev Thirumalai for helpful discussions. We thank Julie Forman-Kay, 
Lewis Kay and Jos\'e Onuchic for their critical reading 
of the manuscript and very helpful comments. This work was supported by 
grant MT-15323 to H.S.C. from the Medical Research Council of Canada.

%*******************************************************************
%\vfill\eject

$\null$\\
$\null$\\

\centerline{\large\bf Appendix}

$\null$ 

\noindent
{\bf Statistical mechanics of a strictly two state model.}\\

Here we describe basic thermodynamics of a strictly 
two-state model, which may be viewed as the $\sH\rightarrow 0$ limit of the 
random-energy model given by Eq.~(7) above. The simplicity of this extreme 
case makes it useful for further elucidating the relationship among different 
midpoint temperatures and van't Hoff enthalpies in the analysis
of calorimetric cooperativity. The strictly two-state model is given by
the partition function
$$
Q(T)= 1+\gD \eHDkT \; ,
\eqno({\rm A}1)
$$
where $\gcD$ in Eq.~(7) is re-written as $\gD$ to highlight that 
we now consider a discrete rather than a continuous density of states.$^{23}$
For this model, $\DHc=\HD$; and the average enthalpy 
$$
\langle H(T) \rangle = {\frac {\gD\HD\eHDkT} {1+\gD\eHDkT}} \; .
\eqno({\rm A}2)
$$
It follows that the specific heat capacity
$$
C_P = {\frac {\partial \langle H(T) \rangle} {\partial T}} 
    = {\frac {{\HD}^2} {\kB T^2}}
      {\frac {\gD \eHDkT} {(1+\gD\eHDkT)^2}} \; .
\eqno({\rm A}3)
$$
This functional form gives a single maximum value for 
$C_P$ at a certain $T=T_{\rm max}$. The relation between
$T_{\rm max}$ and the population midpoint temperature
$$
T_{1/2}={\frac {\HD} {\kB \ln\gD}}\;
\eqno({\rm A}4)
$$
may be determined as follows. First, we note that the slope
of the specific heat function at the population midpoint
$$
{\frac {dC_P} {dT}}\Biggr\vert_{T=T_{1/2}}
= -{\frac {{\HD}^2} {2\kB (T_{1/2})^3}} < 0 \; .
\eqno({\rm A}5)
$$
This establishes $T_{1/2}>T_{\rm max}$ for a strictly two-state
model. We then seek a good estimate of $T_{\rm max}$ by attempting
an approximate solution to the $dC_P/(dT)=0$ condition --- which
is equivalent to the equation
$$
\gD {\rm e}^{-\xi} = {\frac {\xi -2} {\xi +2}} \; ,
\eqno({\rm A}6)
$$
where $\xi=-\HD/(\kB T_{\rm max})$. For $\ln\gD \gg 1$, which is
a reasonable assumption for proteins, as discussed in the text,
$$
T_{\rm max}\approx {\frac {\HD} {\kB [\ln\gD + 4/(2+\ln\gD)]}} < T_{1/2}\; .
\eqno({\rm A}7)
$$
The last inequality follows from Eq.~(A4) for $T_{1/2}$, and confirms
the conclusion we have drawn from Eq.~(A5). Finally, since by Eqs.~(A2) and
(A4) $\langle H(T_{1/2})\rangle$ $=$ $\DHc/2$, we have $T_d=T_{1/2}$. 
Therefore, for a strictly two-state model, 
$$
T_d=T_{1/2}>T_{\rm max} \; .
\eqno({\rm A}8)
$$

We now turn to the various van't Hoff to calorimetric enthalpy ratios
considered in the text [Eq.~(6)]. Obviously, by definition $\kappa_0=1$ 
for the strictly two-state model. Moreover, by Eqs.~(A3) and (A4),
$$
2T_{1/2}\sqrt{\kB C_P(T_{1/2})} = \HD = \DHc \; .
\eqno({\rm A}9)
$$ 
Hence $\kappa_1=\kappa_3=1$ as well, because $T_{1/2}=T_d$. On the other hand,
$$
\kappa_2 = 2T_{\rm max}\sqrt{\kB C_P(T_{\rm max})}/\HD = 
\sqrt{1-4(\kB T_{\rm max}/\HD)^2} < 1 \; .
\eqno({\rm A}10)
$$
However, for proteinlike systems, $\HD \gg \kB T$ is expected 
for any $T$ between 0$^\circ$ to 100$^\circ$C, hence
$T_{1/2}=T_{\rm max}$ and $\kappa_2=1$ are very good approximations.
For instance, if we use the parameters in the text for $\HD$ and $\gD$,
which were motivated by experimental data on CI2 (Fig.~3 of Ref.~54), we 
get $T_{1/2}=336.190$K, whereas $T_{\rm max}=336.025K$ is 
only $0.17^\circ$C lower, and $\kappa_2=0.9997$. Therefore, for a strict 
two-state model with these proteinlike parameters, practically all 
three midpoint temperatures are identical, and all $\kappa$'s are equal 
to one.

%*****************************************************************************

%\par\vfill\eject
%
\vskip 1cm
\noindent
{\large\bf References}

\kern -1.5cm

%\footnotesize

\end{multicols}
%========================================================================

\par\vfill\eject

\centerline{\large \bf Table I}
\vskip .2 in

\begin{center}
\begin{tabular}{|c||c|c|c|c|}
\hline
$T_{\rm midpoint}$ & $\Delta H_{\rm vH}/\Delta H_{\rm cal}$ & references &
           $\Delta H_{\rm vH}/\Delta H_{\rm cal}$ & references \\
\hline
\hline
$T_{1/2}$ & $\kappa_0$ & Ref.~23, Eq.~(4) & & \\
&$\theta =$ [D] &&&\\
\hline
$T_{1/2}$ & $\kappa_1$ & Ref.~23 & $(\kappa_1)^2$ & Ref.~23 \\
\hline
$T_{\rm max}$ & 
$\kappa_2$ & Ref.~40, Eq.~(39) & $(\kappa_2)^2$ 
& Ref.~40, Eq.~(38)\\ 
&&&&Ref.~41, Eq.~(21) \\
\hline
$T_d$ & $\kappa_3$ & Ref.~50, Eq.~(7) & $(\kappa_3)^2$ &
 Ref.~51, Eq.~(11) \\
&&&
$\theta=\langle\Delta H\rangle/\Delta H_{\rm cal}$ &
Ref.~22, Eq.~(22) \\
\hline
\end{tabular}
\end{center}
\vskip .15 in

{\noindent {{\bf Table~I.}}} $\quad$ 
Different definitions in the literature for 
$\Delta H_{\rm vH}/\Delta H_{\rm cal}$,
the van't Hoff to calorimetric enthalpy ratio.
$T_{\rm midpoint}$ is the midpoint temperature of the given definition(s);
see Eq.~(6) in the text. Equation numbers in the table are those in the 
example reference(s) in which a given formula is used or proposed.
$\theta$'s are shown only for $\DHv/\DHc$'s that follow directly from 
Eq.~(4). Note that $\kappa_0$, 
$\kappa_2$, $(\kappa_2)^2$, and $(\kappa_3)^2$ are 
equal, respectively, to the expressions ``$\DHv/\DHc$,'' 
``$\Delta H_{\rm vH}^{\rm exp}/\DHc$,''
``$\Delta H_{\rm vH}^{\rm exp(a)}/\DHc$,'' and
``$\Delta H_{\rm vH}^{\rm exp(a)\prime}/\DHc$'' in Ref.~23.

\vfill\eject

%========================================================================

\centerline{\large \bf Table II}
\vskip .2 in

\begin{center}
\begin{tabular}{|l||c|c|c|c|c|c|}
\hline
\multicolumn{1}{|c||}{Model} & 
\multicolumn{1}{|c|}{$\Delta H_{\rm cal}$} & 
\multicolumn{4}{|c|}{$\Delta H_{\rm vH}/\Delta H_{\rm cal}$} & 
\multicolumn{1}{|c|}{$\Omega_c$} \\ \cline{3-6}
$\null$ & $\null$ & $\kappa_0$ & $\kappa_1$ & $\kappa_2$ & $\kappa_3$ &  \\
\hline
\hline
(a) 2-letter (27mer) &$68.5$& $0.26$ & $0.32$ & $0.36$ & $0.24$ & $11.2$\\
\hline
(b) 3-letter (27mer) &$73.9$& $0.36$ & $0.43$ & $0.46$ & $0.31$ & $20.7$\\
\hline
(c) 20-letter (36mer) &$15.0$& $0.10$ & $0.12$ & $0.67$ & $0.66$ & $38.9$\\
\hline
(d) G\=o (48mer) &$55.2$& $0.54$ & $0.78$ & $0.87$ & $0.87$ & $192$\\
\hline
(e) Modified ``HP'' (36mer) &$35.1$& $0.17$ & $0.23$ & $0.33$ & $0.31$&$12.4$\\
\hline
(f) Sidechain (15mer) &$11.6$& $0.05$ & $0.07$ & $0.38$ & $0.36$ & $5.69$\\
\hline
\end{tabular}
\end{center}
\vskip .15 in

{\noindent {{\bf Table~II.}}} $\quad$ 
Calorimetric cooperativity of the lattice protein models in Figure~3.
Thermodynamic quantities are deduced from Figures~4--9: $\kappa_0$ involves 
the population-based van't Hoff enthalpy,$^{23}$ which can be readily
read off from the $\langle\Delta H\rangle_{\rm D}$ curves. $\kappa_1$, 
$\kappa_2$, and $\kappa_3$ [Eq.~(6)] are deduced from the $C_P$ functions, 
and $\Delta H_{\rm cal}$ is obtained by numerical integration of $C_P$ 
over $T$. The Klimov-Thirumalai$^{48}$ cooperativity parameter $\Omega_c$ 
is calculated for these models and included for comparison; the present
$\Omega_c=5.69$ is slightly different from the value $5.32$ reported
by Klimov and Thirumalai.$^{48}$

\vfill\eject

%========================================================================
\centerline{\large \bf Table III}
\vskip .2 in

\begin{center}
\begin{tabular}{|l||c|c|c|c|c|c|}
\hline
Model & $T_{\rm max}$ & $C_{P,{\rm max}}$ & $C_{P,{\rm max}}^{\rm (s)}$ &
$\Delta H_{\rm vH}^{\rm (s)}$&$\Delta H_{\rm cal}^{\rm (s)}$
&$\kappa_2^{\rm (s)}$\\
\hline
\hline
(a) 2-letter (27mer) & $1.35$ & $80.6$ & $69.5$ & $22.6$ & $24.2$ & $0.932$\\
\hline
(b) 3-letter (27mer) & $1.56$ & $117$ & $105$ & $32.0$ & $33.6$ & $0.952$\\
\hline
(c) 20-letter (36mer) & $0.282$ & $316$ & $294$ & $9.66$ & $10.3$ & $0.943$\\
\hline
(d) G\=o (48mer) & $0.764$ & $986$ & $965$ & $47.5$ & $47.3$ & $1.00$\\
\hline
(e) Modified ``HP'' (36mer) & $0.558$ & $107$ & $102$ & $11.3$&$27.8$&$0.406$\\
\hline
(f) Sidechain (15mer) & $0.268$ & $66.4$ & $59.9$ & $4.14$ &$7.75$&$0.535$\\
\hline
\end{tabular}
\end{center}
\vskip .15 in

{\noindent {{\bf Table~III.}}} $\quad$ 
Effects of baseline subtractions on the predicted
calorimetric cooperativities of the six lattice protein models considered
in this work: The effective van't Hoff to calorimetric enthalpy ratio
$\kappa_2^{\rm (s)}$ (right column) is equal to
$\Delta H_{\rm vH}^{\rm (s)}/\Delta H_{\rm cal}^{\rm (s)}$ [Eq.~(9)]. 
The definitions of all quantities tabulated and methods to determine 
them are described in the text, Figure~11, and upper panels of Figures~12 
and 13. 

\vfill\eject 
%
%========================================================================
\begin{multicols}{2}
\centerline{\large \bf Figure Captions} \vskip .2 in 

{\noindent {\bf Fig.~1}}$\quad$ 
Densities of states $g(H)$ of random energy models.
Each parabolic curve is $\ln g(H)$ from Eq.~(7)
with $\HD =3\times 10^4$ (vertical dashed lines), $\gcD=5.68\times 10^{38}$, 
as described in the text, and $H$ is in units of $\kB$. The $\kappa_0$ 
values of these curves, $0.6$, $0.80$, $0.95$, and $0.98$, quantify
the different degrees of cooperativity of four models given
here as examples, with standard deviations of denatured enthalpy 
$\sigma_H=$ $1800$, $1350$, $700$, and $440\kB$ respectively. 
$\kappa_0$'s are the population-based$^{23}$ $\DHv/\DHc$ ratios.
The horizontal dashed line highlights the fact that for these models it is 
possible for $g(H)<1$; and the dot indicates that their unique native (N) 
states have zero enthalpy [$\delta$-function in Eq.~(7)].
Note that the logarithmic scale along the vertical axis implies that
a 0.693 decrease in $\ln g$ is equivalent to halving the value of $g$
itself. Hence the distribution of $g$ is much sharper than 
this logarithmic plot might have otherwise conveyed.
\vskip .2 in 

{\noindent {\bf Fig.~2}}$\quad$ 
Relationship among different calorimetric two-state criteria in the
random energy models defined by Eq.~(7). See text and Table~I for definitions 
and references. Left column: (a) Midpoint transition temperatures and 
(b) van't Hoff to calorimetric enthalpy ratios, as functions of the standard 
deviation $\sigma_H$ of denatured enthalpy distribution. (b) shows $\kappa$'s 
vs. $\sigma_H$ times a constant, so that the horizontal scale corresponds
to Onuchic et al.'s expression$^{9}$ for $T_{\rm g}/T_{\rm f}$. We note 
that $\kappa_0$ in (b) is well approximated by Eq.~(13) of Ref.~23. Right 
column: Experimental formulas for $\DHv/\DHc$ vs. the population-based 
$\kappa_0$ used in our theoretical analyses.
\vskip .2 in 

{\noindent {\bf Fig.~3}}$\quad$ 
Recent three-dimensional cubic lattice protein models considered in
this paper for their conformities to the calorimetric two-state 
criterion. Monomers (residues) are numbered from one end of the chain
to the other; monomer 1 corresponds to the leftmost letter of a sequence.
Each model protein chain is shown in its unique native or ground-state
(lowest-enthalpy)
structure. The corresponding sequence is also included, except for the G\=o 
model in (d), as the interactions of a G\=o model is determined solely
by the ground-state conformation it presumes. 
(a) A 2-letter model of Socci and Onuchic (sequence 002 in 
Table~1 of Ref.~44). 
(b) A 3-letter model of Socci et al. (sequence in Fig.~3 of Ref.~45).
(c) A 20-letter model of Gutin et al. (sequence in Fig.~1 of Ref.~49).
(d) A G\=o model of Pande and Rokhsar (structure in Fig.~1 of Ref.~46).
(e) A modified HP ``solvation'' model of Sorenson and Head-Gordon (sequence
6 in Table~1 of Ref.~47). Filled and open circles represent the H and P 
monomers, respectively, in this modified HP model.
(f) A 20-letter sidechain model of Klimov and Thirumalai (sequence A in
Fig.~1 of Ref.~48). Here the main-chain monomers are numbered, and sidechains
are represented by grey circles.
\vskip .2 in 

{\noindent {\bf Fig.~4}}$\quad$ 
Thermodynamic cooperativity of the 2-letter model in Fig.~3a. Results are 
obtained by the Monte Carlo (MC) histogram technique using simulation at 
$T=1.5$. [N] and [D] are respectively the fractional native 
and denatured population, [N] $+$ [D] $=1$. In this figure and subsequent 
Figs.~5--9, the native state of each model is taken to be only its single 
ground-state (lowest $H$) conformation, and the denatured state consists of 
all other conformations.$^{23}$ The vertical lines give the midpoint 
temperatures.  From left to right, they are 
$T_{1/2}$ when [N] $=$ [D] $=1/2$ (dashed line), $T_{\rm max}$, and $T_d$ 
(solid lines). In all six models studied here (Figs.~4--9), 
$T_{1/2}<T_{\rm max}<T_d$.
{\it Upper panel}: the specific heat capacity $C_P$ is defined by Eq.~(2) in 
the text; $(C_P)_{\rm D}$ is the specific heat capacity of the denatured
ensemble, obtained by replacing the Boltzmann averages $\langle\dots\rangle$ 
in Eq.~(2) over the full ensemble by averages $\langle\dots\rangle_{\rm D}$ 
over the denatured (nonnative) ensemble.$^{23}$ 
{\it Lower panel}: The excess heat function $\langle\Delta H\rangle$ 
(solid curve increasing with $T$) is given by Eq.~(1) in the text,
$\langle\Delta H\rangle_{\rm D}$ (dashed curve) is the corresponding average 
over the denatured ensemble,$^{23}$ both are normalized by (in units of) 
$\Delta H_{\rm cal}$ obtained by numerical integration of the entire area 
under the $C_P$ curve, part of which is shown in the upper panel.
Our results for $C_P$ and $\langle \Delta H \rangle$ are
numerically consistent with the $C_V$ and $\langle E \rangle$ functions 
in Figs.~10 and 9 of the original study.$^{72}$
\vskip .2 in 

{\noindent {\bf Fig.~5}}$\quad$ 
Same as Fig.~4, but for the 3-letter model in Fig.~3b;
obtained by the MC histogram technique from simulation at $T=1.5$.
\vskip .2 in 

{\noindent {\bf Fig.~6}}$\quad$ 
Same as Fig.~4, but for the 20-letter model in Fig.~3c;
obtained by the MC histogram technique from simulation at $T=0.27$.
\vskip .2 in 

{\noindent {\bf Fig.~7}}$\quad$ 
Same as Fig.~4, but for the G\=o model in Fig.~3d; all continuous
curves are obtained by the MC histogram technique from simulation at
$T=0.75$. For this model, $T_{\rm max}$ ($0.764$) is almost equal to 
$T_d$ ($0.767$). Black dots in the lower panel are fractional native
populations [N] at six different temperatures computed by direct MC 
simulations, showing good agreement with results from the histogram method.
\vskip .2 in 

{\noindent {\bf Fig.~8}}$\quad$ 
Same as Fig.~4, but for the modified HP ``solvation'' model in Fig.~3e;
obtained by the MC histogram technique from simulation at $T=0.6$.
Our simulated $C_P$ function (upper panel) is consistent with the
original simulation ($C_V$ of sequence 6 in Fig.~8 of Ref.~47).
\vskip .2 in 

{\noindent {\bf Fig.~9}}$\quad$ 
Same as Fig.~4, but for the 20-letter sidechain model in Fig.~3f;
obtained by the MC histogram technique from simulation at $T=0.25$.
The $C_P$ function in the upper panel is consistent with 
the original heat capacity simulation ($C_V$ in Fig.~2c of Ref.~48).
Our results are also consistent with the thermodynamics properties 
$\langle\chi\rangle$, $\Delta\chi$, and $P_{\rm NBA}$ given by Klimov 
and Thirumalai$^{48}$ in their Fig.~2 (data not shown).
\vskip .2 in 

{\noindent {\bf Fig.~10}}$\quad$ 
Distributions of denatured (nonnative) enthalpy $H$ of the 48mer G\=o
model in Figs.~3d and 7 at different temperatures $T$, obtained by
direct MC simulations (same temperatures as the black dots in Fig.~3).
The native enthalpy is $-57$. The total area under a distribution curve 
is proportional to the fractional denatured population [D] at the given 
temperature.
\vskip .2 in 

{\noindent {\bf Fig.~11}}$\quad$ 
Exploring effects of baseline subtractions on predicted calorimetric 
cooperativity. {\it Ad hoc} baseline subtractions are applied to the 
heat capacity functions 
of the 2-letter (a), G\=o (b), modified HP (c), and 20-letter sidechain (d)
models. The model heat capacities ($C_P$'s) are the same as those presented 
in Figures~4 and 7--9. In each plot, the shaded area is subtracted from 
the original (pre-subtraction) $\Delta H_{\rm cal}$ to yield a new
effective calorimetric enthalpy $\Delta H_{\rm cal}^{\rm (s)}$ 
($<\Delta H_{\rm cal}$). Native and denatured baselines with non-zero slopes
are constructed for (b) and (d). Denatured baselines with negative slopes 
are provided for (a) and (c), but their native baselines are assumed to 
have zero slope (i.e., no new native baseline) because the 
significant curvatures of their $C_P$ 
functions at low temperatures do not appear to warrant linear positive-slope 
extrapolations. Solid vertical lines mark the temperature $T_{\rm max}$ 
at the peak of heat capacity functions; the black dot marks the arithmetric 
mean of the values of native and denatured baselines at $T_{\rm max}$. 
Following standard experimental calorimetric baseline procedures$^{50,51}$ 
(see also Ref.~22), the new effective heat capacity peak value 
$C_{P,{\rm max}}^{\rm (s)}$ is given by the vertical measure between the 
black dot and the pre-subtraction $C_{P,{\rm max}}=C_P(T_{\rm max})$.
The quantities $C_{P,{\rm max}}^{\rm (s)}$ and 
$\Delta H_{\rm cal}^{\rm (s)}$ are then used to compute the new effective 
van't Hoff to calorimetric enthalpy ratios $\kappa_2^{\rm (s)}$ in Table~III.
Included for comparison are nonlinear ``formal two-state'' baselines 
(dotted curves) constructed using the method of Zhou et al.$^{22}$ 
Nonlinear baselines correspond to heat capacity functions $(C_P)_0$ and
$(C_P)_1$ of the native and denatured ensembles respectively. No 
native nonlinear baseline is provided for (a) -- (c)
because each of their native states is taken to have only a single
conformation, as in the original analyses.$^{44,46,47}$ Hence $(C_P)_0=0$ 
and $(C_P)_1=(C_P)_{\rm D}$ for (a) -- (c). On the other hand,
for the 20-letter sidechain model in (d), the nonlinear native baseline is 
calculated$^{22}$ from a multiple-conformation native state 
defined by the original authors.$^{48}$ 
Vertical dashed lines mark the temperature $T_m$. For (a) -- (c),
$T_m=T_{1/2}$; for (d), $T_m$ is the temperature at which one half of 
the chain population is in the multiple-conformation native state (``native
basin of attraction'') defined in Ref.~48. See the text for further details.
\vskip .2 in 

{\noindent {\bf Fig.~12}}$\quad$ 
Thermodynamic/calorimetric cooperativity of a 3-letter model.
(a) Same as Figure~11a, but for the 3-letter model of Socci et al.$^{45}$
in Figures~3b and 5. (b) Root-mean-square radius of gyration $R_g$ of this
3-letter chain model vs. temperature. (Square root of the Boltzmann average
of square radius of gyration of the chains.) $R_g$ continues to increase
substantially as temperature is raised well above the transition region 
(vertical dashed and solid lines).
\vskip .2 in 

{\noindent {\bf Fig.~13}}$\quad$ 
Thermodynamic/calorimetric cooperativity of a 20-letter model.
{\it Upper panel}: Same as Figure~11d, but for the 20-letter model of
Gutin et al.$^{49}$ in Figures~3c and 6. As in Figure~11d, the vertical
dashed line marks the temperature $T_m$ at which one half of the
chain population is in the multiple-conformation native state defined
by the original authors as the ensemble of conformations that have more than
20 contacts that also occur in the ground-state conformation
(${\bf Q}>20$, {\bf Q} is referred to as the number of native
contacts).$^{49}$ For this 36mer model, the total number ${\bf Q}_{\rm N}$
of native contacts equals 40. The corresponding native and denatured
nonlinear baselines are calculated using the method of Zhou et al.$^{22}$  
{\it Lower panel}: Folding/denaturation transition tracked by different
order parameters. [N] is the fractional chain population 
in the single-conformation
ground state; $\langle{\bf Q}\rangle/{\bf Q}_{\rm N}$ is the normalized 
Boltzmann-averaged number of native contacts; $P({\bf Q}>20)$ is the 
fractional population in the multiple-conformation native state; and
$\langle\chi\rangle$ is the Boltzmann average of the overlap function 
$\chi$ of Thirumalai and coworkers,$^{48}$ which is a useful measure
of the structural similarity between any given conformation and the 
ground-state conformation.  The single ground-state conformation have
${\bf Q}/{\bf Q}_{\rm N}=1$ and $\chi=0$. The {\it inset} in the upper
panel shows the relation between ${\bf Q}/{\bf Q}_{\rm N}$ and $\chi$. 
While each ${\bf Q}$ is consistent 
with many values of $\chi$, 
and vice versa (scatter plot), for this model the correlation between their 
{\it Boltzmann averages} at different temperatures is almost perfect (curve 
in inset with slope $\approx -1$).
\vskip .2 in 

{\noindent {\bf Fig.~14}}$\quad$ 
Conformational diversity in the multiple-lattice-conformation
native state of the 20-letter model in Figures~3c, 6 and 13.
In each conformation, the directionality of the sequence is
indicated by the filled circle, which marks the position of monomer 1 
in Figure~3c. The three rows show example non-ground-state conformations
({\it from top to bottom}) with number of native contacts ${\bf Q}=34$, 
$35$, and $36$ respectively. These {\bf Q} values are close to the average 
${\bf Q}$ of the multiple-conformation native state at the midpoint
temperatures $T_m$ and $T_{\rm max}$ (vertical dashed and solid lines in 
Figure~13).$^{49}$ 
\vskip .2 in 

{\noindent {\bf Fig.~15}}$\quad$ 
Effects of the folding/denaturation transition on conformational properties 
of the 20-letter model in Figures~3c, 6 and 13. The dashed lines on the left 
mark $T_{1/2}$ at which the fractional population [N] of the single 
ground-state conformation equals $1/2$, the dashed lines on the right mark 
$T_m$ $\approx$ $T_{\rm max}$ (see Figure~13). {\it Upper panel}: 
Boltzmann-averaged number of nonnative contacts (i.e., contacts that do not 
belong to the single ground-state conformation) vs. temperature.
{\it Lower panel}: Root-mean-square radius of gyration vs. temperature.
(Same as Figure~12b, but now for the 20-letter model.)
\vskip .2 in 
\end{multicols}
\vfill\eject
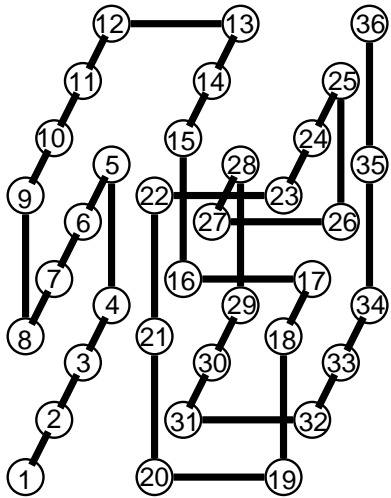
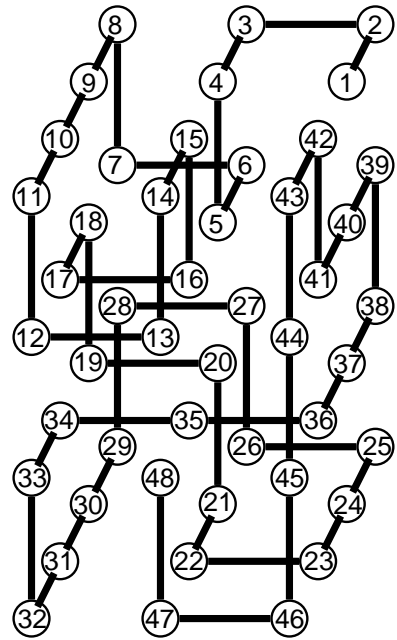
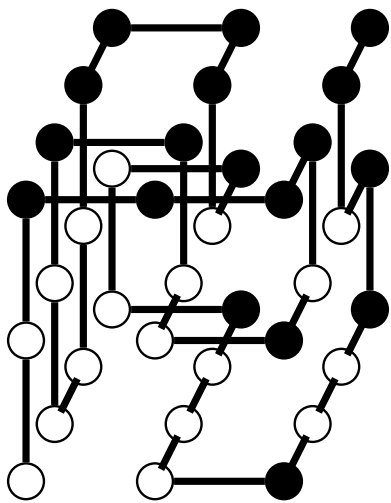
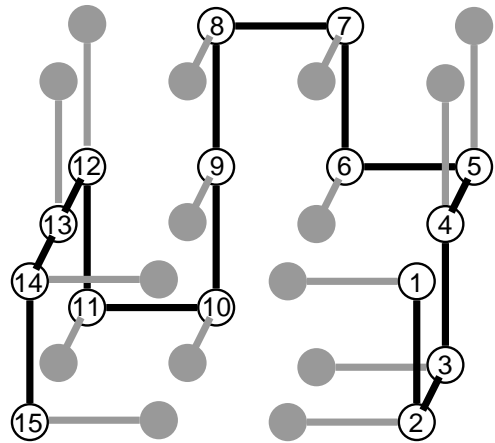
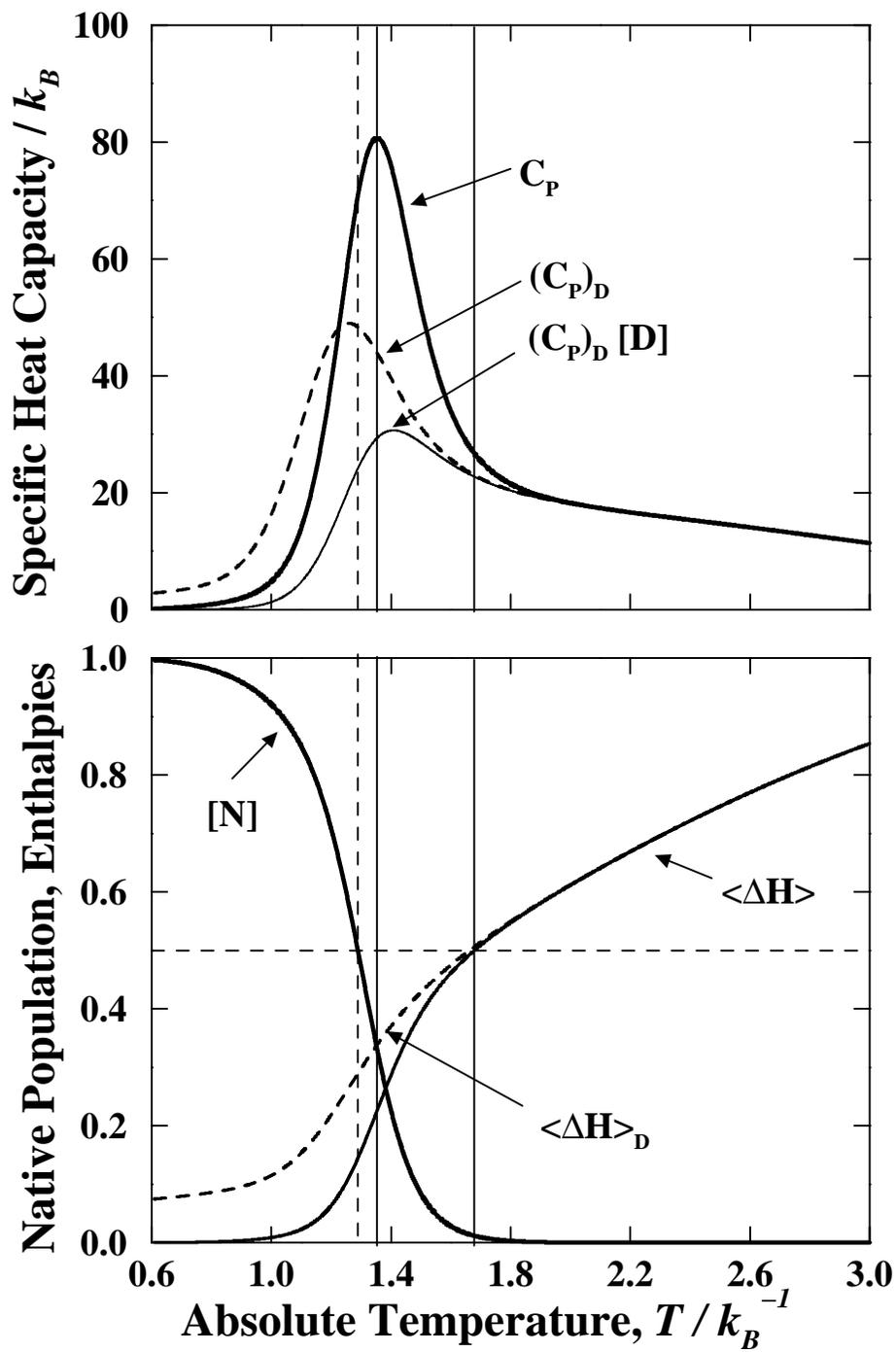
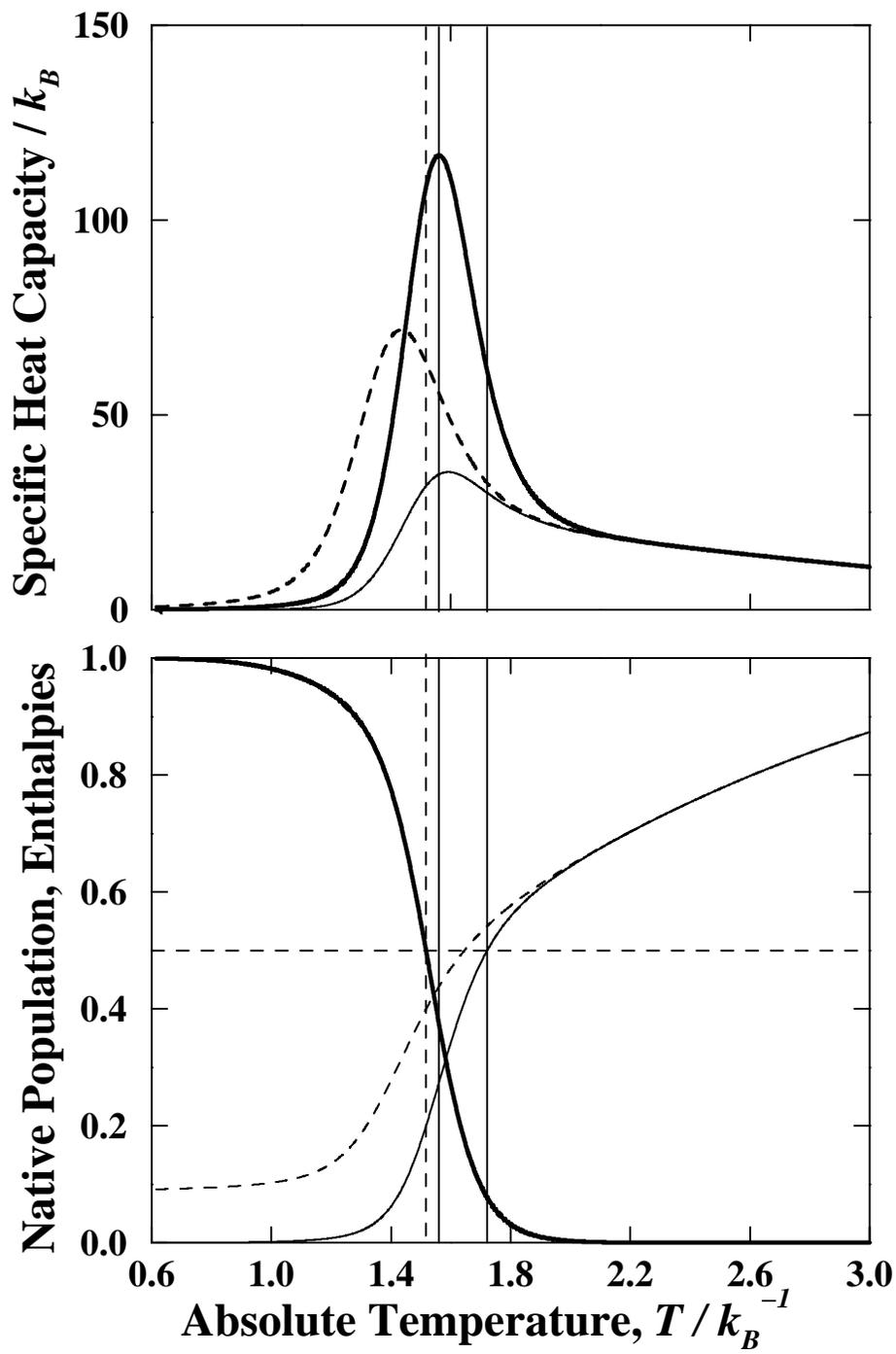
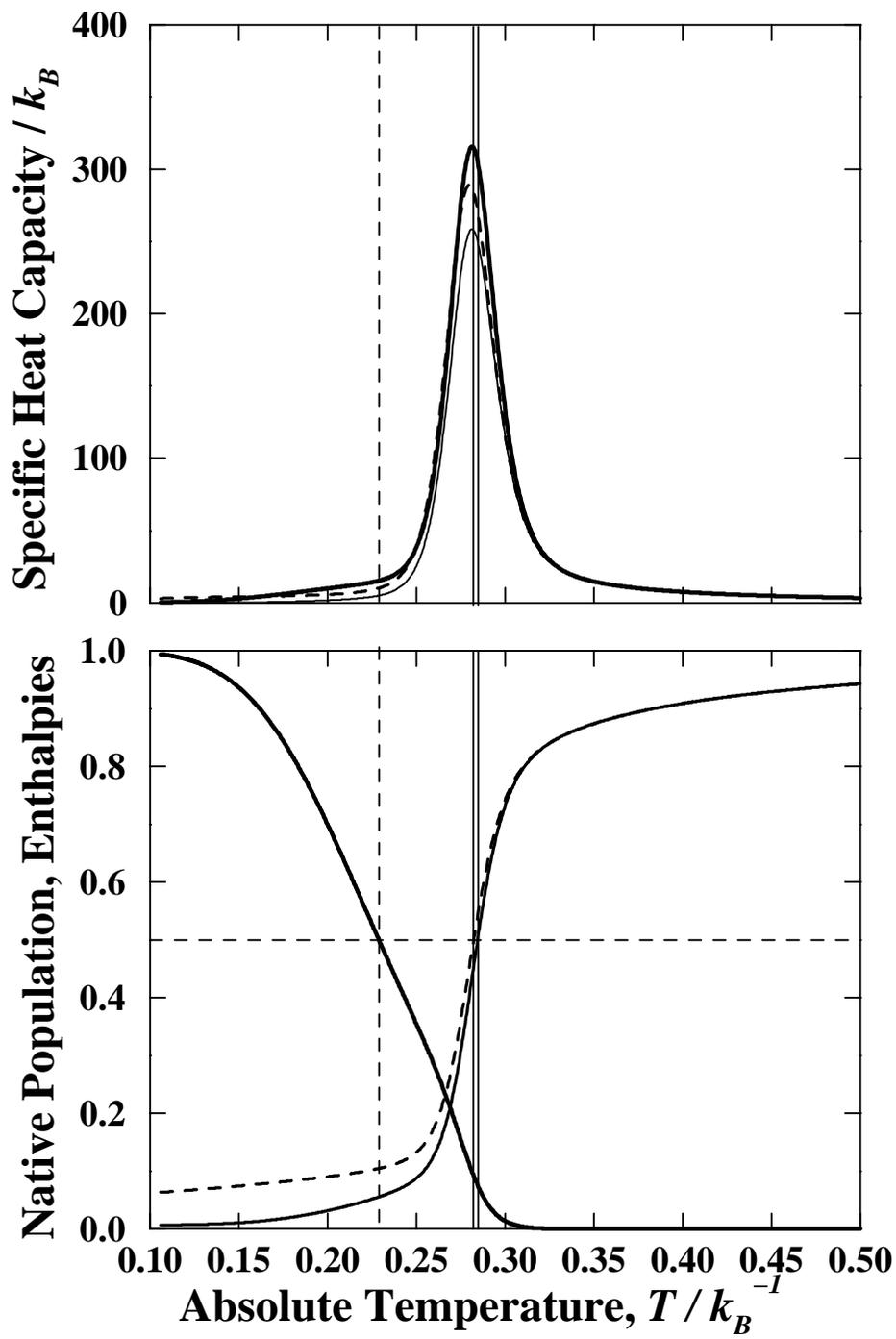
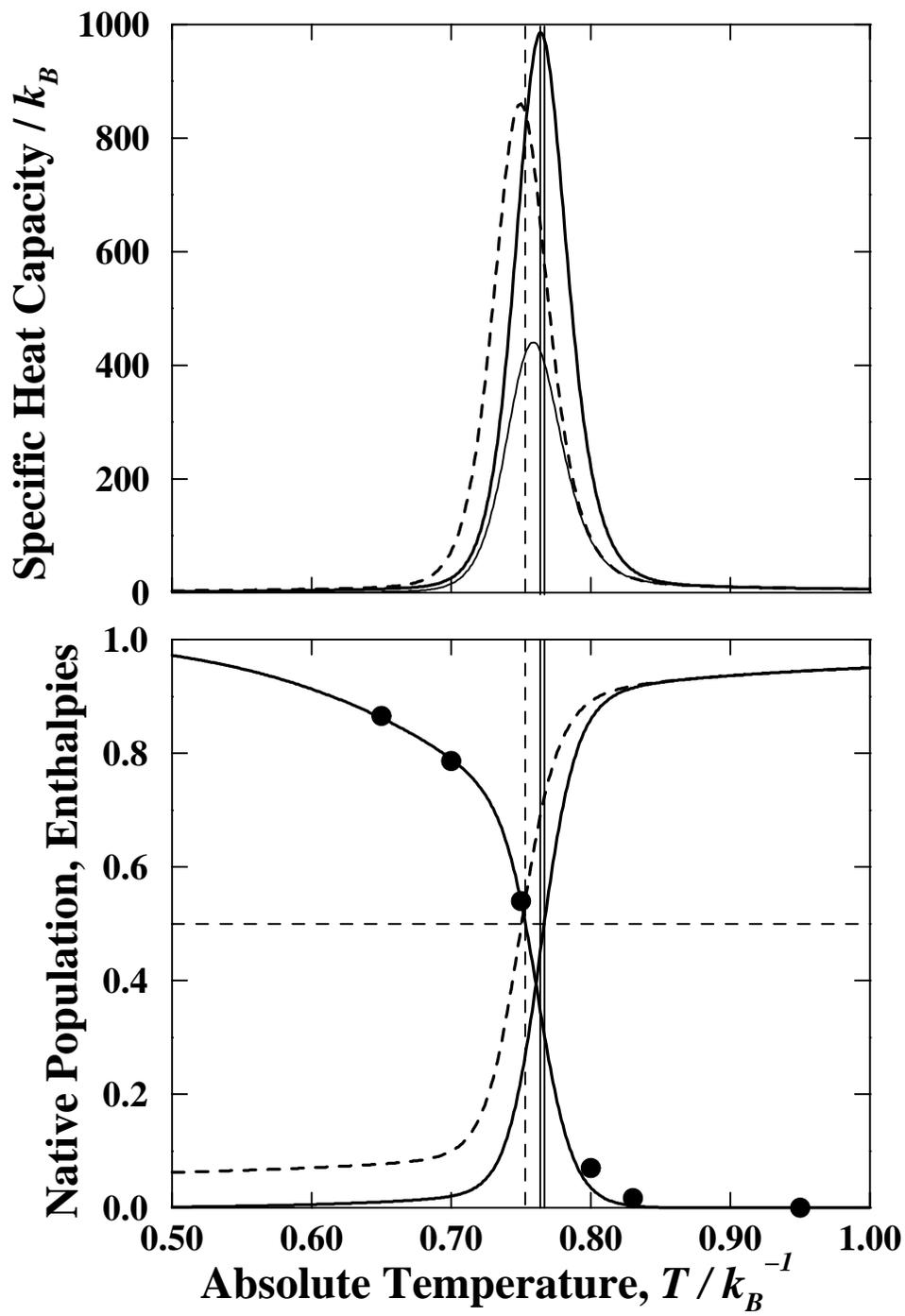
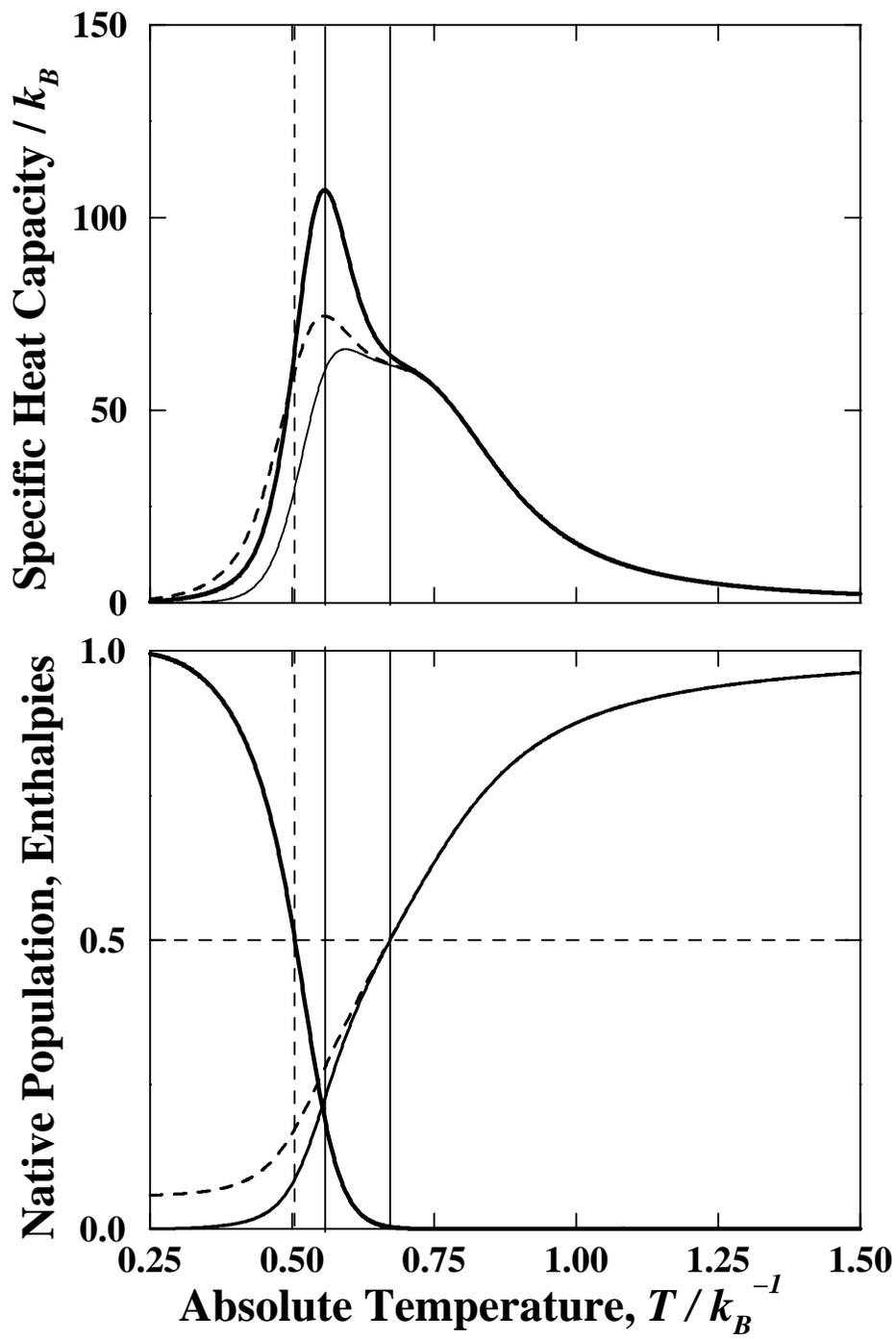
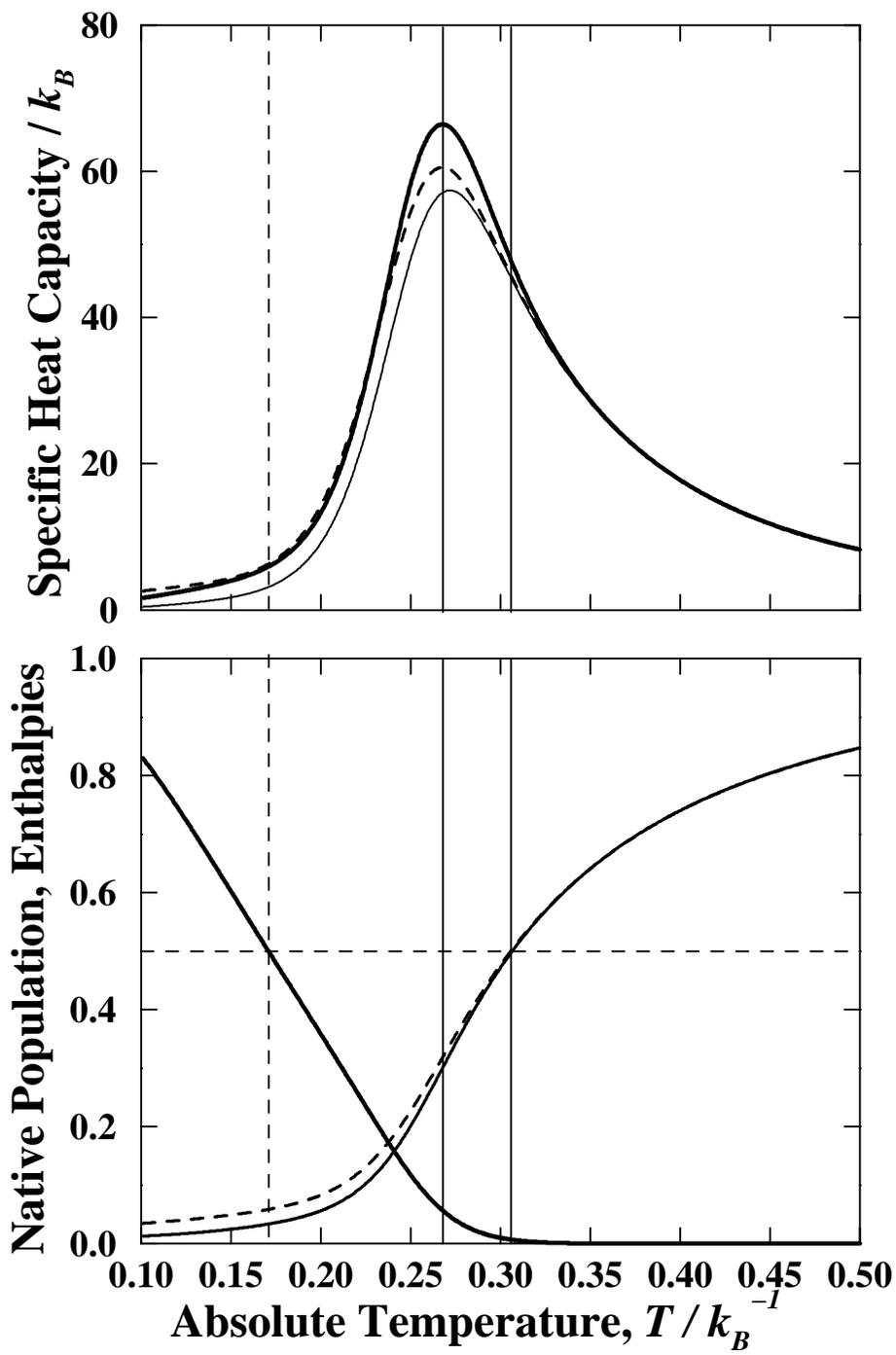
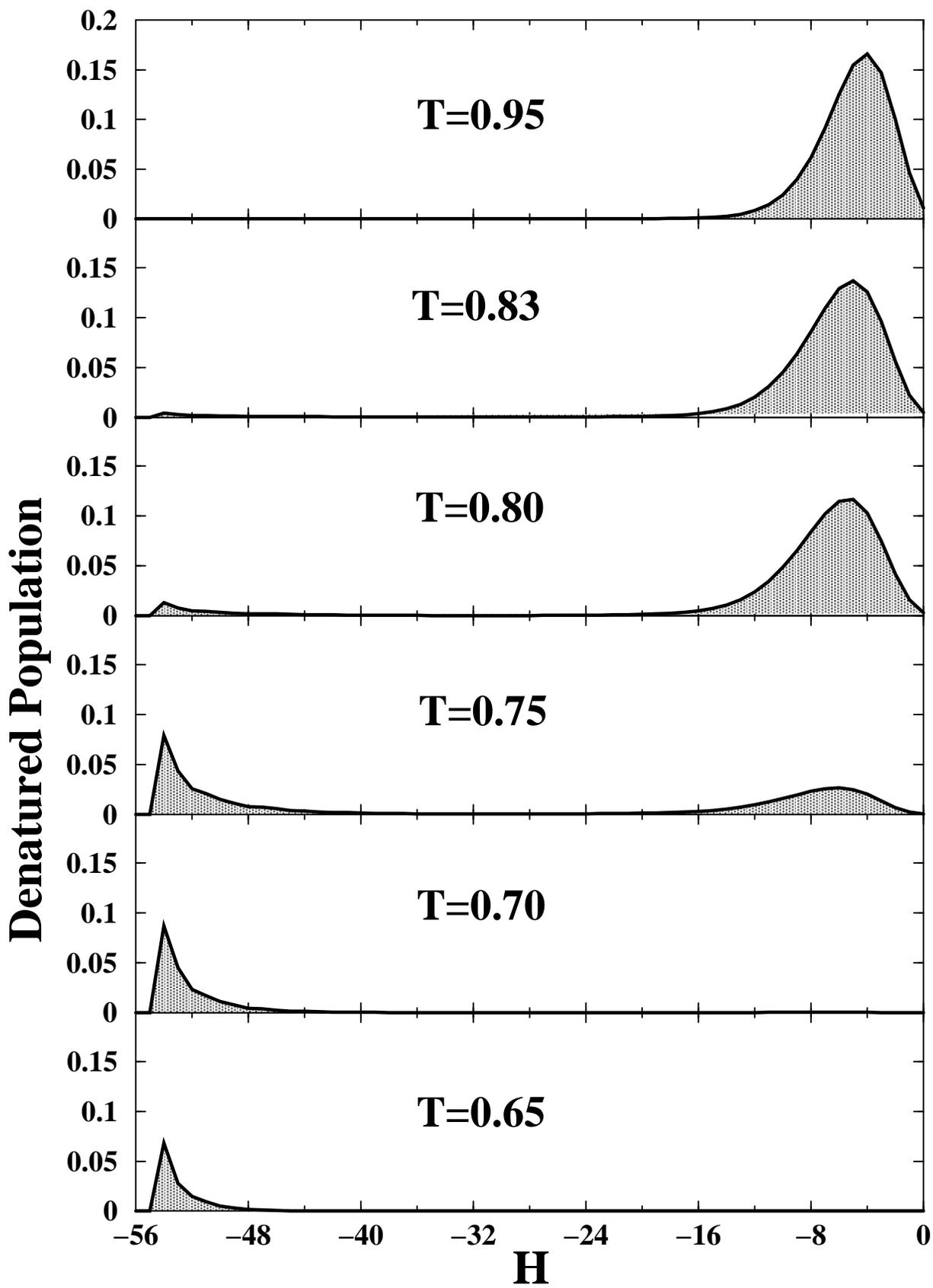
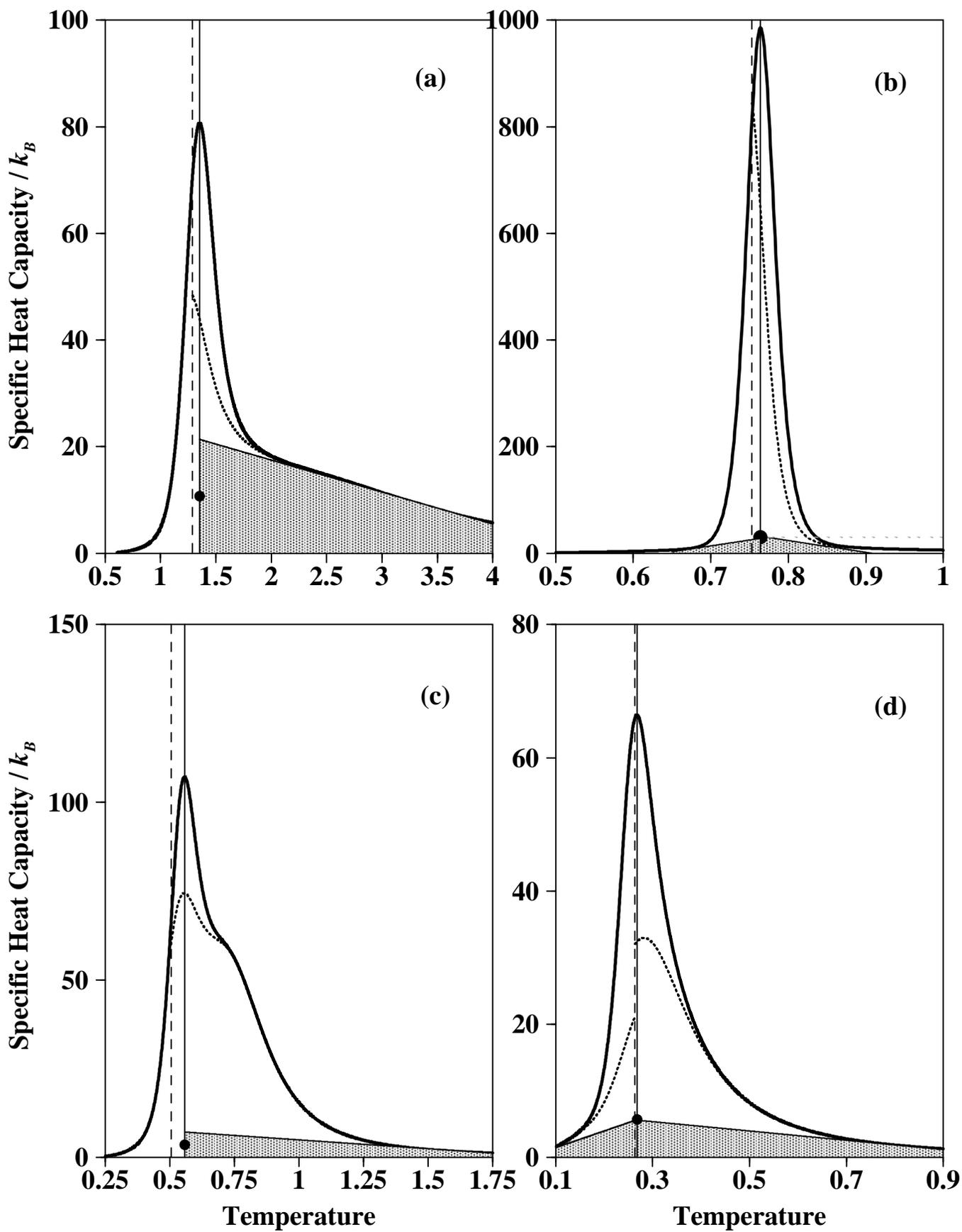
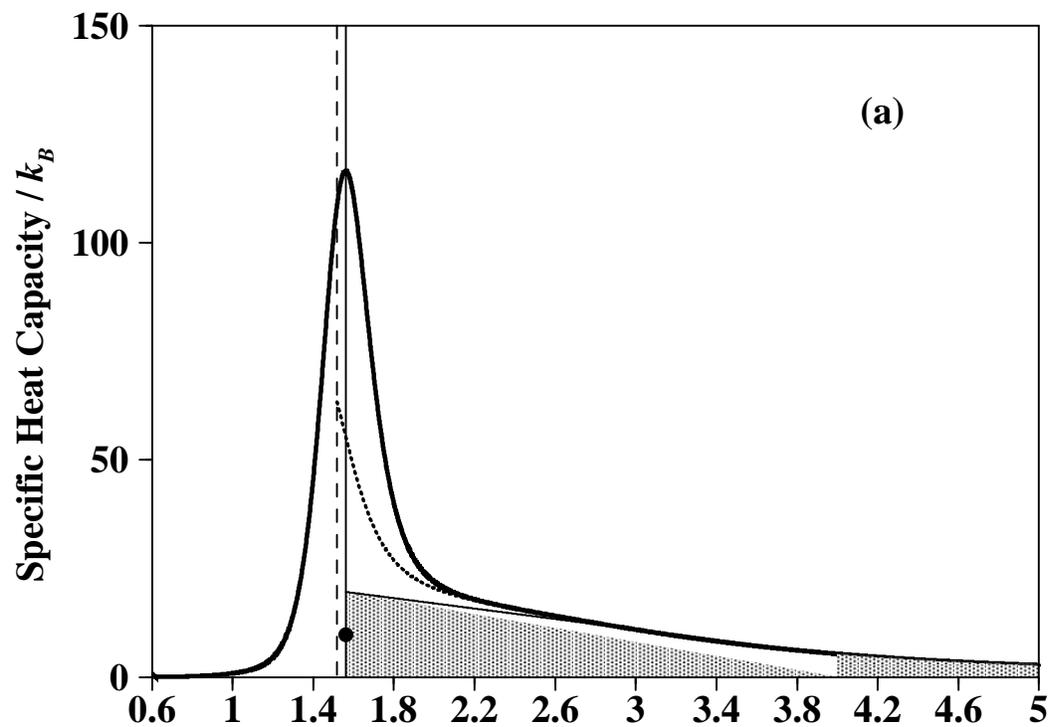
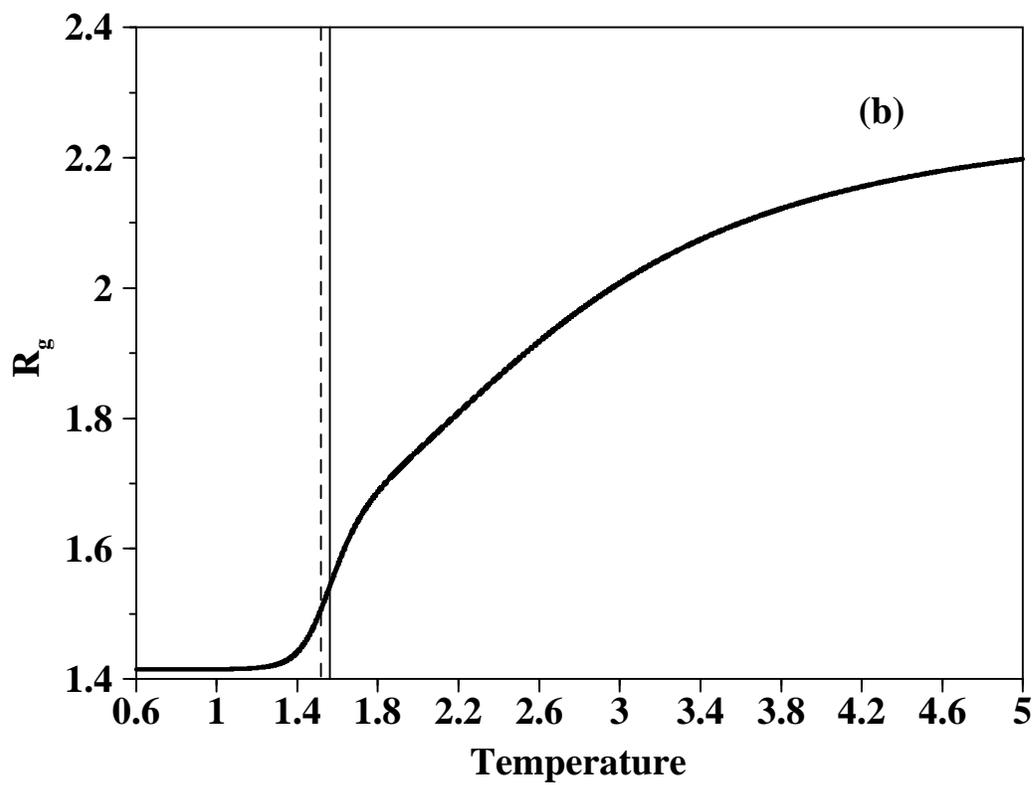
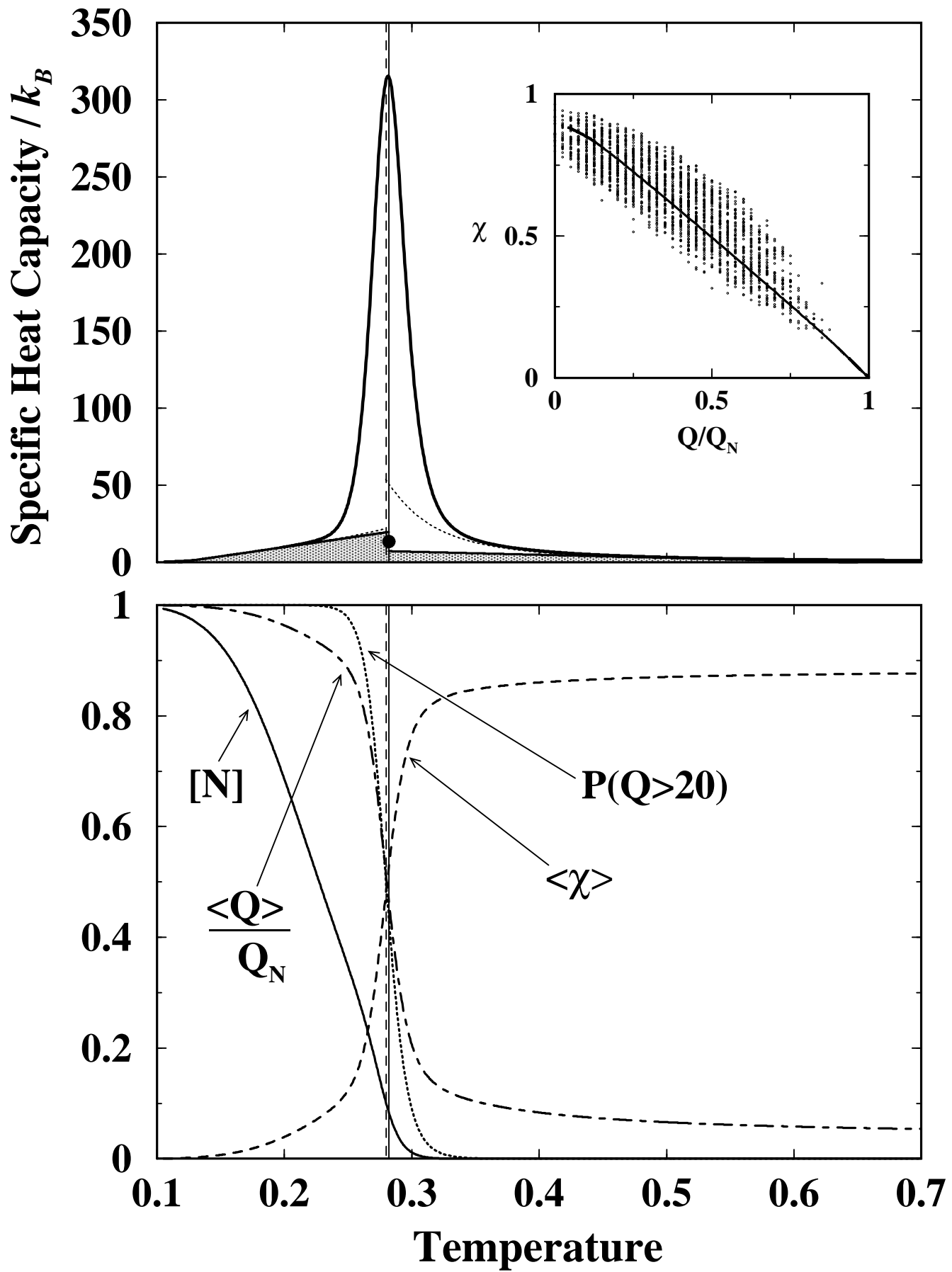
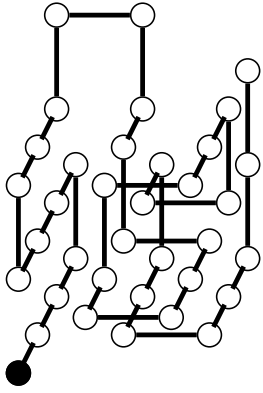
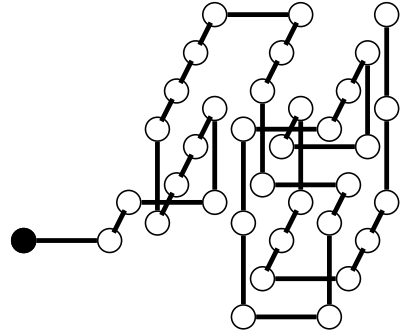
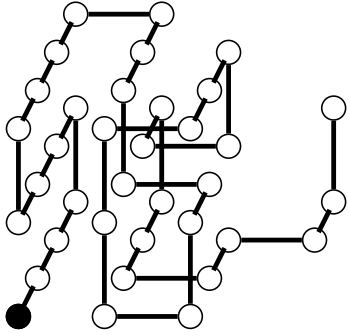
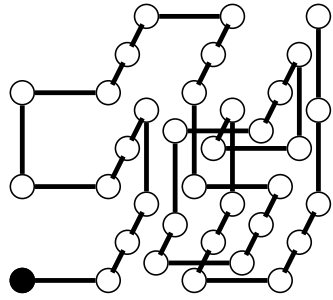
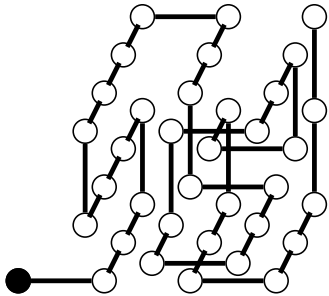
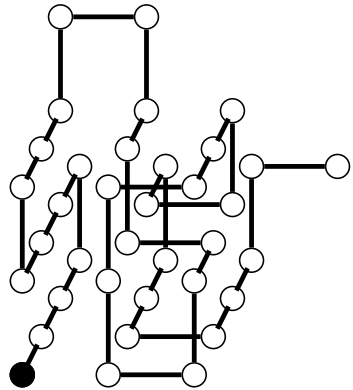
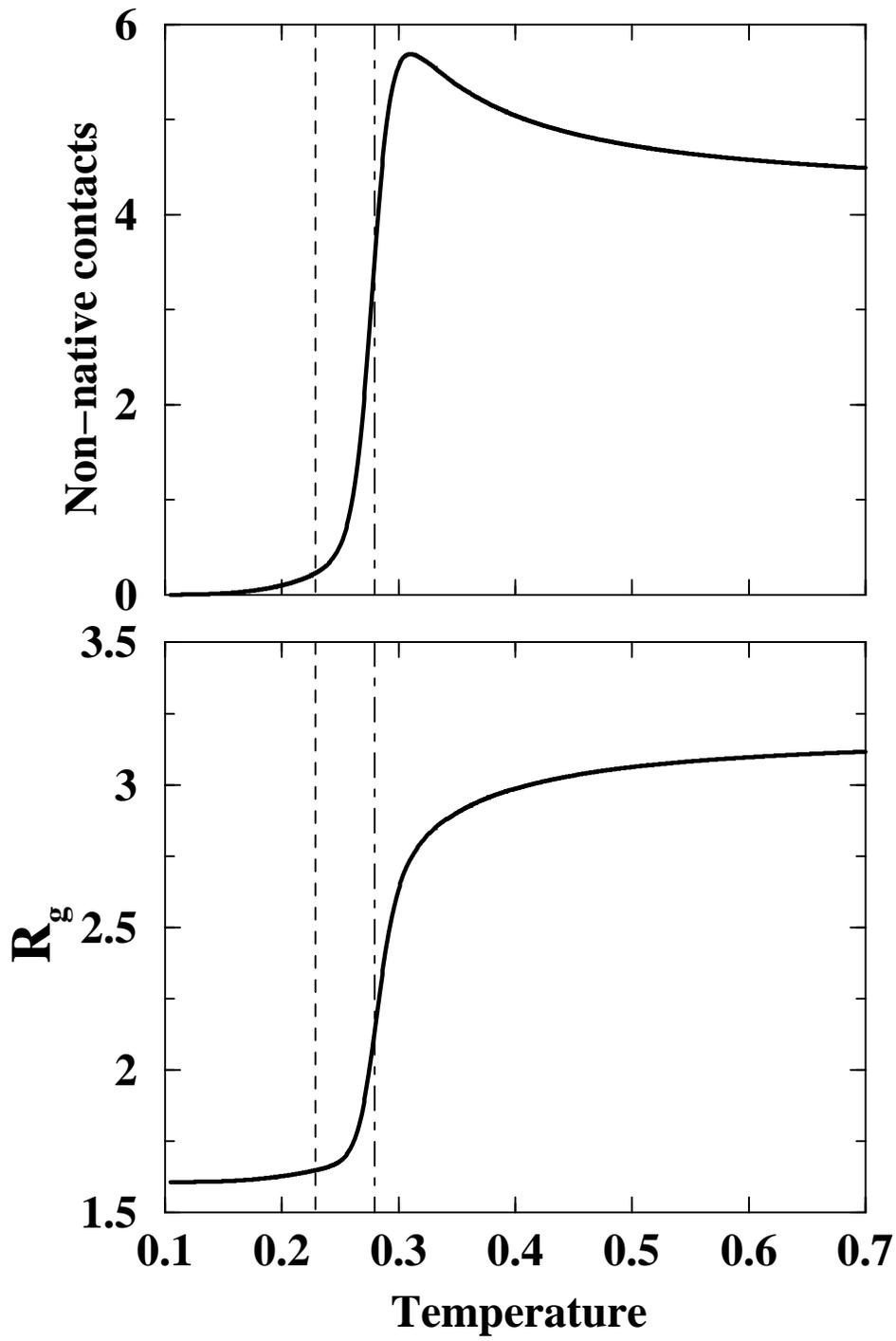
\end{document}